\documentstyle[11pt,aas2pp4,psfig]{article}

\newcommand{\as}{\mbox{$''$}}
\newcommand{\etal}{{\it et~al.\/}}
\newcommand{\Hline}[1]{\mbox{H{\footnotesize {#1}}}}
\newcommand{\Halpha}{\Hline{\mbox{$\alpha$}}}
\newcommand{\Hbeta}{\Hline{\mbox{$\beta$}}}
\newcommand{\HI}{\mbox {H\thinspace{\footnotesize I}}}
\newcommand{\HII}{\mbox {H\thinspace{\footnotesize II}}}
\newcommand{\kms}{\mbox{km\thinspace s$^{-1}$}}
\newcommand{\Msun}{\mbox{${\cal M}_\odot$}}

\begin{document}

\title{The Panchromatic Starburst Intensity Limit\\
At Low And High Redshift \footnote{Based on observations with the
NASA/ESA {\em Hubble Space Telescope\/} obtained at the Space
Telescope Science Institute, which is operated by the Association of
Universities for Research in Astronomy, Inc., under NASA contract
NAS5-26555.}}

\author{Gerhardt R.\ Meurer, Timothy M.\ Heckman}
\affil{The Johns Hopkins University, Department of Physics and
Astronomy,\\ Baltimore, MD 21218-2686\\
Electronic mail: meurer@pha.jhu.edu,heckman@pha.jhu.edu}

\author{Matthew D. Lehnert}
\affil{Sterrewacht Leiden, Postbus 9513, 2300 RA, Leiden, The
Netherlands.\\ Electronic mail: lehnert@strw.LeidenUniv.nl}

\author{Claus Leitherer} 
\affil{Space Telescope Science Institute,
3700 San Martin Drive, Baltimore, MD 21218 \\ Electronic mail:
leitherer@stsci.edu}

\author{James Lowenthal\altaffilmark{2}}
\affil{Lick Observatory, University of California Santa Cruz\\
Electronic mail: james@lick.ucsc.edu}

%
% pp version
%
\vspace{1cm}
\centerline{Accepted for publication in the {\em Astronomical Journal\/}.}
\vspace{1cm}
\centerline{Received: \underline{~ 10 March, 1997 ~~~}
Accepted: \underline{~ 9 April, 1997 ~~~}}

\altaffiltext{2}{Hubble Fellow.}

\clearpage % pp version only

\begin{abstract} 

The integrated bolometric effective surface brightness $S_e$
distributions of starbursts are investigated for samples observed in
1.\ the rest frame ultraviolet (UV), 2.\ the far-infrared and \Halpha,
and 3.\ 21cm radio continuum emission.  For the UV sample we exploit a
tight empirical relationship between UV reddening and extinction to recover
the bolometric flux.  Parameterizing the $S_e$ upper limit by the 90th
percentile of the distribution, we find a mean $S_{e,90} = 2.0 \times
10^{11}\, L_\odot\, {\rm kpc^{-2}}$ for the three samples, with a
factor of three difference between the samples.  This is consistent with
what is expected from the calibration uncertainties alone.  We find
little variation in $S_{e,90}$ with effective radii for $R_e \sim 0.1
- 10$ kpc, and little evolution out to redshifts $z \approx 3$.  The
lack of a strong dependence of $S_{e,90}$ on wavelength, and its
consistency with the pressure measured in strong galactic winds, argue
that it corresponds to a {\em global\/} star formation intensity limit
($\dot\Sigma_{e,90} \sim 45\, {\cal M}_\odot\, {\rm kpc^{-2}\,
yr^{-1}}$) rather than being an opacity effect.  There are several
important implications of these results: 1.\ There is a robust
physical mechanism limiting starburst intensity. We note that
starbursts have $S_e$ consistent with the expectations of
gravitational instability models applied to the solid body rotation
portion of galaxies.  2.\ Elliptical galaxies and spiral bulges can
plausibly be built with maximum intensity bursts, while normal spiral
disks can not.  3.\ The UV extinction of high-$z$ galaxies is
significant, implying that star formation in the early universe is
moderately obscured.  After correcting for extinction, the observed
metal production rate at $z \sim 3$ agrees well with independent
estimates made for the epoch of elliptical galaxy formation.

\end{abstract}


\section{Introduction\label{s:intro}}

Starbursts are regions of intense massive star formation that can
totally dominate a galaxy's integrated spectrum.  They range in size
from giant extragalactic \HII\ regions of scale size $\sim 0.1$ kpc
(e.g.\ NGC~604 in M~33) to global bursts (often in merging systems)
many kpc across that can cover the entire face of the system (e.g.\
NGC~4038/4039, NGC~4449).  There has been much interest in the
starburst phenomenon over the past decade or so, yet there is much
more to learn.  For example we do not know how starbursts are
regulated: what turns them on and off, and what keeps them going.
Also unknown is how the properties of starbursts vary with size or
luminosity.  We address some of these issues in this paper. 

While starbursts are worthy of investigation in their own right, they
are even more important when placed in the broader context of
contemporary extragalactic astrophysics.  The cosmological relevance
of starbursts has been underscored by the recent discovery of: the
existence of a population of high-redshift ($z > 2$) UV-bright field
galaxies (cf. Steidel \etal\ 1996a,b; Lowenthal \etal\ 1997). The
sheer number density of these galaxies implies that they almost
certainly represent precursors of typical present-day galaxies in an
early actively-star-forming phase. This discovery moves the study of
the star-forming history of the universe into the arena of direct
observations (cf. Madau \etal\ 1996), and gives added impetus to the
quest to understand local starbursts. In particular, a thorough
understanding of how to exploit the diagnostic power of the rest-frame
ultraviolet (UV) properties of local starbursts will give astronomers
powerful tools with which to study star-formation in the early
universe.

Accordingly, we have been using the HST to determine the basic UV
structure and morphology of starbursts. In our first paper (Meurer
\etal\ 1995; hereafter M95) we analyzed in detail UV HST images of nine
starburst systems with distances\footnote{We convert all results to
$H_0 = 50\, {\rm km\, s^{-1}\, Mpc^{-1}}, q_0 = 0$, except where
noted.} $D < 75$ Mpc. Here, we turn our gaze outward and consider the
UV structure of starbursts from $z = 0$ to $z > 3$, in an attempt to
document and understand the sytematic properties of starbursts.  We
also broaden our net by considering the structural properties of
infrared-selected ``dusty'' starbursts.

M95 showed that nearby starbursts are irregular structures consisting
of diffusely distributed light interspersed with prominent compact
(radii $\leq 10$ pc) star clusters. Numerous other HST studies have
also commented on the presence of luminous young star clusters
(including Holtzman \etal, 1992; Whitmore \etal, 1993; Conti \&\
Vacca, 1994; Hunter \etal, 1994; Bower \&\ Wilson, 1995; O'Connell
\etal, 1995; Whitmore \& Schweizer, 1995; Holtzman \etal, 1996; Maoz,
\etal\ 1996a,b; Schweizer \etal, 1996; Watson \etal, 1996).  While the
clusters are certainly the most striking aspect of the images, it is
the diffuse light that dominates.  M95 find on average 80\%\ of the UV
light is diffusely distributed, and 20\%\ comes from clusters.  Maoz
\etal\ (1996a) find similar fractional UV contributions of clusters in
five starburst ring galaxies.  M95 argue that there are two modes of
star formation in starbursts: prominent cluster formation, and
dominant diffusely distributed star formation.  

M95 also noted that the effective surface brightnesses of most of the
starbursts in their (small) sample span a narrow range of values
(dispersion of 0.4 mag arcsec$^{-2}$ for 8/11 starburst regions).
Since starbursts produce the most intense UV emission observed in
galaxies, this result implies that there is an upper limit to the UV
surface brightness of galaxies.  Lehnert \&\ Heckman (1996) found a
similar limit to the far-infrared surface brightness of far-infrared
galaxies (FIRGs).  In both samples the illuminating source can be
traced to high-mass stars.  The UV light studied by M95 is dominated
by the light from stars with mass $\sim$ 20 \Msun.  The far-infrared
emission studied by Lehnert \&\ Heckman is thought to result from dust
absorbing UV -- optical photons (which are predominantly produced by
high-mass stars) being heated and reemitting the radiation thermally
at infrared wavelengths.  Thus FIRGs are thought to be obscured
starbursts.  These surface brightness limits suggest a limit to the
intensity of high-mass star formation, which implies that there is a
mechanism that limits the global star formation intensity of galaxies.

Here we reexamine the issue of starburst intensities.  Since selection
effects are strongest at low surface brightness, we are primarily
interested in the intensity maximum.  Our aims are to determine what
the starburst intensity limit is, investigate whether it varies with
certain parameters, and consider what causes the limit.  By comparing
starbursts selected and observed at different wavelengths we address
the issue of opacity and the difference between dusty and relatively
dust free starbursts.  By comparing starbursts at different redshifts
we are able to consider the evolution of starburst intensities, and
the resultant cosmological implications.  Our method is to compile
observations of a diverse set of starburst samples from the
literature, combined with (a dash of) new data.  After defining some
relevant quantities in \S\ref{s:quant} we compile three samples of starburst
data: I.\ a sample of starbursts observed in the rest-frame vacuum
ultraviolet (\S\ref{s:uvsamp}); II.\ a sample observed in the
far-infrared and \Halpha\ (\S\ref{s:oirsamp}); and III.\ a sample
observed at radio wavelengths (\S\ref{s:radsamp}).  In
\S\ref{s:caveats} we discuss the selection effects and measurement
biases, and estimate the level of agreement we expect between
intrinsically similar samples.  In \S\ref{s:disc} we discuss the
interpretation of the results and their implications, and
\S\ref{s:conc} summarizes the conclusions.

\section{Definition of Quantities}\label{s:quant}

To quantify the intensity of high-mass star formation we measure the
surface brightness at a wavelength that traces high-mass stars, within
an aperture that encompasses half of the {\em total\/} emission from
the starburst.  Thus we employ the effective, or half light, radius
$R_e$, and effective surface brightness $S_e$ enclosed within $R_e$.
These quantities are best measured with a curve of growth extracted
from a background-subtracted image of the starburst.  Relative to the
total luminosity $L$, the effective surface brightness is given by:
\begin{equation}
S_e = \frac{L}{2\pi R_e^2}. \label{e:sedef}
\end{equation}
Note that the definition of surface brightness used by Lehnert \&\
Heckman (1996) is a factor of two higher than our definition.  Their
definition is equivalent to saying that all of the starburst is
located within the half light radius of the tracer.  Here we assume a
one-to-one spatial relationship between the high-mass stars and their
trace emission.  One complication is that the apertures used to
measure $R_e$ are often elliptical, not circular, in shape.  Our
approach is to take $R_e$ as the geometric equivalent radius of the
aperture:
\begin{equation}
R_e =  a_e \sqrt{(b/a)_e}
\end{equation}
where $a_e$ is the semimajor axis length of the elliptical aperture
enclosing half the the total flux, and $(b/a)_e$ is the corresponding
axial ratio at $a_e$.

We emphasize that $R_e$ and $S_e$ are global quantities; that is they
are associated with the total light of the starburst.  This includes
both clusters and diffuse light. Furthermore the area over which the
$S_e$ is measured scales with the distribution of the light. For
moderate to high redshift galaxies full cosmological corrections have
been applied to the $R_e$ and $L$ measurements assuming an open
universe with $H_0 = 50\, {\rm km\, s^{-1}\, Mpc^{-1}}$ and $q_0 = 0$.

The salient properties of starbursts are essentially defined by
ionizing stars.  These have masses $\gtrsim 20~M_\odot$, and hence
lifetimes $\lesssim 10$ Myr.  Ideally we would like to determine how
these stars are distributed.  However, the size of the ionizing source
distribution is not directly observable since either very little
ionizing emission escapes from starbursts, or it is absorbed by the
local ISM/IGM before reaching the earth (Leitherer \etal, 1996).
Therefore $R_e$ of some other tracer of the ionizing population is
employed: the vacuum-UV (sample I), \Halpha\ (sample II), and 21cm
continuum radio emission (sample III).  To compare results we convert
all flux related measurements to bolometric quantities.  The
algorithms we use are given below.

To characterize the $S_e$ distribution of a sample we consider the
sample median and 90th percentiles, which we denote as $S_{e,50}$ and
$S_{e,90}$, respectively.  The samples were compiled from a
hodge-podge of studies with selection criteria and measurement
techniques varying from study to study.  In addition, there is a well
known bias when observing extended objects to observe those with the
highest surface brightness (e.g.\ Disney 1976; McGaugh \etal\ 1995),
so the lower percentiles of the $S_e$ distributions are likely to be
rather incomplete.  This is especially true for starbursts which are
often recognized by their high surface brightness.  Therefore,
$S_{e,50}$ and $S_{e,90}$ should not be interpreted too literally.
They actually represent upper limits to the true median and 90th
percentiles of the $S_e$ distributions of star forming galaxies.  Here
$S_{e,50}$ should be interpreted as the typical surface brightness
selected in a sample, and $S_{e,90}$ should be interpreted as close
to the highest surface brightness found by a given technique.  In
\S\ref{s:caveats} we discuss the selection effects and biases of the
samples and estimate the level of agreement we expect.

\section{Ultraviolet sample}\label{s:uvsamp}

\subsection{Ultraviolet extinction}\label{ss:dust}

There are many appealing reasons to observe starbursts in the
vacuum-UV (see M95).  However, there is one serious hindrance to
interpreting the results: dust.  It efficiently extincts and scatters
UV radiation.  When dust absorbs UV and optical photons it heats up
and reemits the radiation thermally in the far-infrared.  The
redistribution of radiant energy must be well modeled and corrected
for in order to determine the true bolometric luminosity of a
starburst.

\begin{figure}
\centerline{\hbox{\psfig{figure=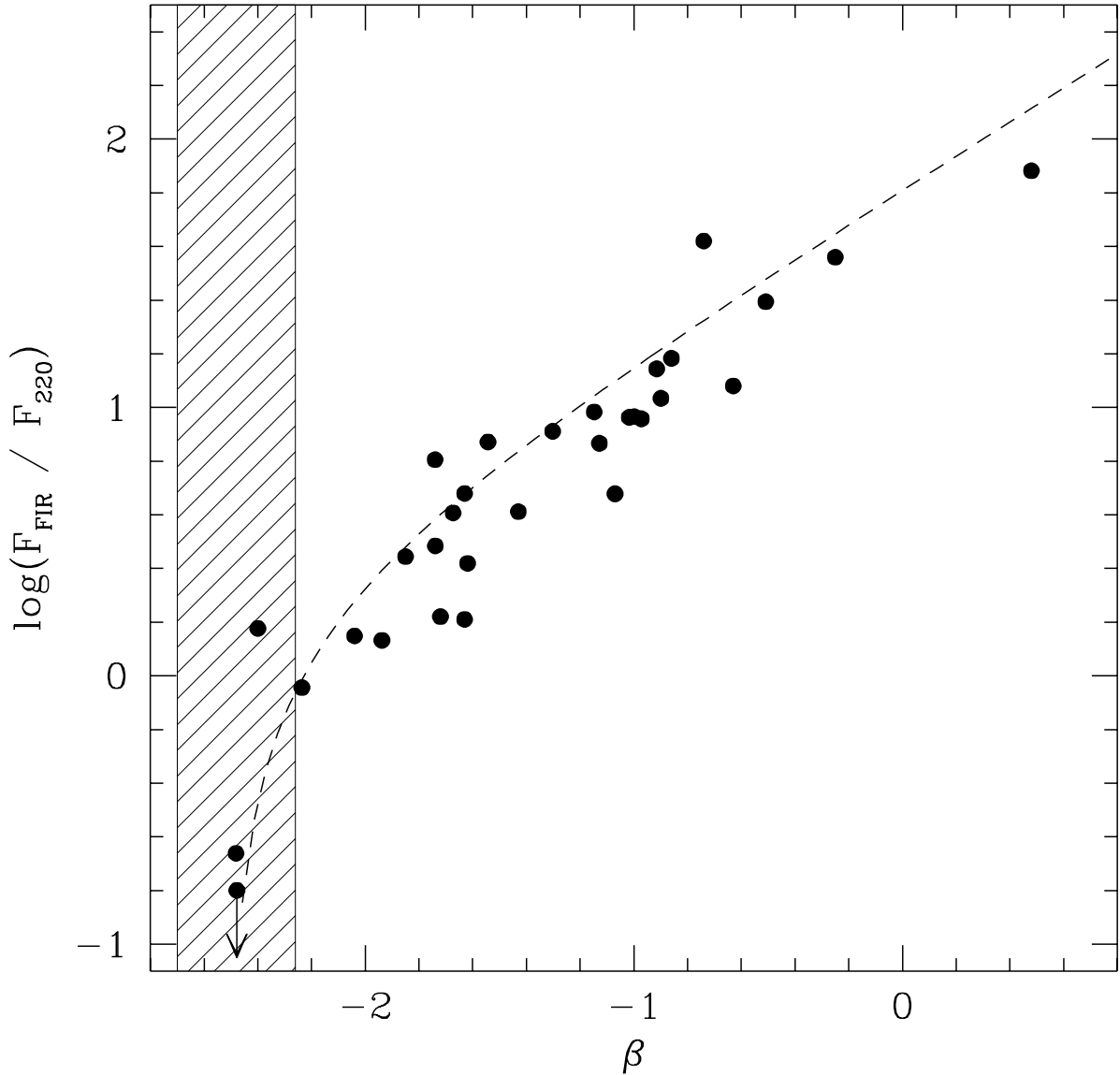,width=17cm}}}
\caption{The ratio of far-infrared to UV fluxes
compared to ultraviolet spectral slope $\beta$, adapted from Fig.~6 of
M95.  The data correspond to UV-selected starbursts observed by IRAS
and HST or IUE.  Only galaxies with isophotal diameters $D_{25} \leq 5'$
are plotted so as to exclude galaxies with large $F_{220}$ aperture
corrections.  The hatched region shows the expected $\beta$ for
naked ionizing populations.  The dashed line shows the expected
relationship for a starburst having an intrinsic $\beta_0 = -2.5$ that
is reddened and extincted by a foreground screen of dust with the Calzetti
\etal\ (1994) extinction law. \label{f:uvir}}
\end{figure}

M95 showed that it is possible to do just that using only the UV
properties of starbursts.  This is illustrated in Fig.~\ref{f:uvir}
which is adapted from Fig.~6 of M95.  It shows the ratio of
far-infrared (FIR) flux $F_{\rm FIR}$ to UV flux $F_{220}$ compared to
the spectral slope $\beta$ which is a measure of the ultraviolet
color, for a sample of UV-selected starbursts.  The definitions of
these quantities are given below (\S\ref{s:irmeth}, \S\ref{s:uvmeth})
and in M95.  In this plot the $y$ axis quantifies the redistribution
of spectral energy from the UV to the FIR.  A strong relationship is
apparent in this diagram - in the sense that, as relatively more light
is emitted in the FIR, the starbursts become redder.  Note that the
hatched region in this plot shows the expected range of colors for
naked ionizing populations.  All the galaxies in this plot display
strong recombination emission spectra, hence their intrinsic colors
should be that of an ionizing population.  Figure~\ref{f:uvir} then
indicates that reddening is positively correlated with dust
extinction.  Such a correlation is a major prediction of simple
foreground screen models for dust extinction.  The dashed line in
Fig.~\ref{f:uvir} shows the expected relationship for a starburst
having $\beta = -2.5$ and foreground screen dust having the Calzetti
\etal\ (1994; hereafter C94) ``extinction law''\footnote{We use the
term ``extinction law'' rather loosely.  The C94 law is better
referred to as an {\em attenuation law} since it is based on
observations which recover most or all of the UV flux, hence
extinction is partially compensated for by foreward scattering, thus
producing a relatively grey attenuation law.}.  Details of the model
can be found in M95.

Figure~\ref{f:uvir} shows that a foreground screen dust geometry
models well the redistribution of spectral energy from the UV to the
FIR.  Similar results have been noted by Calzetti \etal\ (1996) and
Lehnert \&\ Heckman (1996).  However, dust in the vicinity of
starbursts does not necessarily have such a simple distribution in all
cases.  The rms in $F_{\rm FIR}/F_{220}$ about the model line of $\sim
0.4$ dex is probably larger than the errors, and may be indicative of
the non-uniform nature of the dust distribution (e.g.\ dust lanes).
In addition, Fig.~\ref{f:uvir} plots only UV-selected starbursts,
which preferentially will have little extinction.

The point is that the tight empirical relation between UV color and
infrared to UV flux ratio shown in Fig.~\ref{f:uvir} gives us a method
to remove the effects of dust extinction for UV-selected starbursts.
We adopt the C94 curve in Fig.~\ref{f:uvir} because it is physically
plausible and goes through the data fairly well.  Whether or not the
details of this model are absolutely correct is irrelevant, as long as
this empirical relationship holds for all UV-selected starbursts.
Other plausible fits through the data could also be used, and they
will yield the same UV correction factor as we derive to within 0.4
dex.

\subsection{Calculation of intrinsic quantities}\label{s:uvmeth}

In all cases measurements of $L$, $R_e$, and thus $S_e$ have been made
from HST images obtained in the rest frame UV.  In order to make
extinction and $k$ corrections it is crucial to measure the UV color,
or equivalently the UV spectral slope, since the observed UV continuum
spectra of starbursts are well fit by a power law of index $\beta$:
\begin{equation}
f_\lambda \propto \lambda^\beta \label{e:beta}
\end{equation}
(C94, M95).  Here $f_\lambda$ is the spectral flux density per
wavelength interval. In order to compare the subsamples we reference
all observations to the observations of M95.  These employed the {\em
Faint Object Camera\/} with F220W filter, which has central wavelength
$\lambda_c = 2320$\AA.  The total flux is given as $F_{220} =
\lambda_c f_{220}$.  The methods used to estimate $\beta$ vary from
subsample to subsample, and are described below.

The first step in converting the UV to a bolometric flux is to apply
(where appropriate) the $k$ correction:
\begin{equation}\label{e:kcorr}
k(220) = \frac{f_{2320}}{f_{\lambda_c/(1+z)}}
  = \left\{ \frac{(1+z)2320{\rm \AA}}{\lambda_c} \right\}^\beta.
\end{equation}
Here $\lambda_c$ corresponds to the central wavelength of the filters
used below.  After correcting for Galactic foreground extinction, the
UV fluxes were then corrected for intrinsic extinction using the C94
model line in Fig.~\ref{f:uvir} as discussed above.  This model yields
the intrinsic extinction $A_{\rm int}$ following the algorithm in M95.
For this model a change in spectral slope $\Delta\beta = 0.1$
corresponds to $\Delta A_{\rm int} \approx 0.2$ mag for small $\beta -
\beta_0$.

\begin{figure}
\centerline{\hbox{\psfig{figure=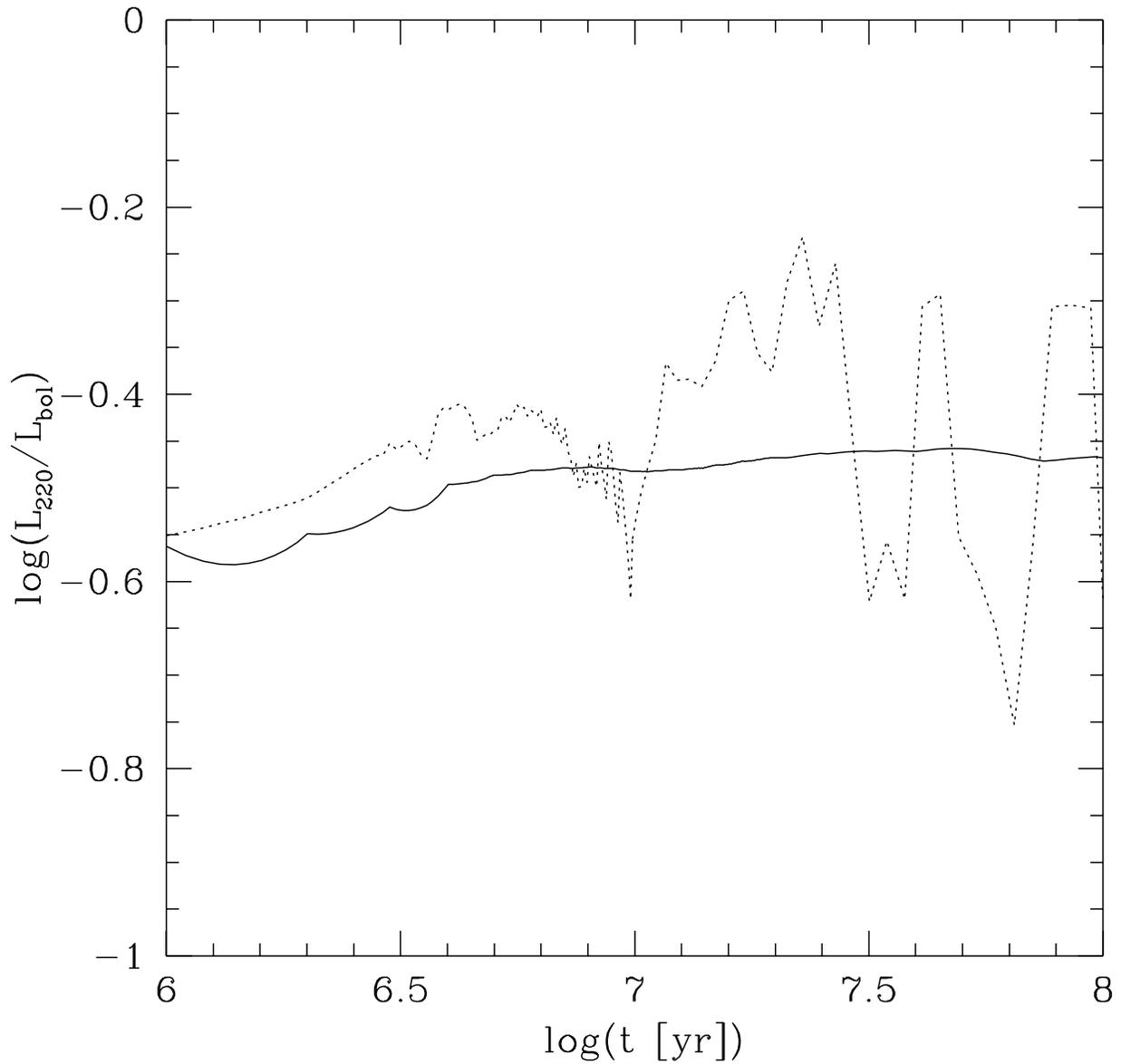,width=17cm}}}
\caption{The temporal evolution of F220W luminosity
as a fraction of the bolometric luminosity for a naked starburst
population.  The curves were derived from the models of Leitherer and
Heckman (1995), assuming a Salpeter (1955) IMF slope, between mass
limits of 1 and 100 \Msun.  The effect of extending the mass range
down to 0.1 \Msun\ is negligible. The solid line is for a constant
star formation rate population, of duration $t$, and the dotted line
is for an instantaneous starburst with age $t$. \label{f:bcev}}
\end{figure}

We use the population models of Leitherer \&\ Heckman (1995; hereafter
LH95) to convert from intrinsic F220W flux/luminosity to bolometric
flux.  We assume that starbursts are well represented by a solar
metallicity 10 Myr duration constant star formation rate population
having a Salpeter (1955) initial mass function (IMF) spanning the mass
range of 0.1 to 100 \Msun. The 10 Myr duration is meant to represent
the typical crossing time of a starburst as estimated by M95.
Although longer durations are likely in the largest starbursts, the UV
properties will be dominated by the stars less than 10 Myr old.  In
any case the choice of a particular age has only a minor effect on the
determination of bolometric quantities. This is demonstrated in
Fig.~\ref{f:bcev} which shows the evolution of $L_{220}/L_{\rm bol}$
at solar metallicity for both instantaneous burst and constant star
formation rate populations for this IMF slope (see M95 and LH95 to see
the evolution of the numerator and denominator separately).  For our
adopted population model,
\begin{equation}
\frac{L_{220}}{L_{\rm bol}} = 0.33. \label{e:bcuv}
\end{equation}
Figure~\ref{f:bcev} shows that for this IMF slope the adopted
bolometric correction is accurate to about 20\%\ or better for
instantaneous burst populations less than 10 Myr old (i.e.,\ ionizing
bursts) and constant star formation rate populations over the full
age range shown.  For our adopted starburst population parameters the
mass to light ratio is
\begin{equation}
\frac{\cal M}{L_{\rm bol}} = 2.2 \times 10^{-3}~ \frac{{\cal
M}_\odot}{L_\odot}, 
\label{e:mlrat}
\end{equation}
and the relationship between star formation rate $\dot{{\cal M}}$ and $L_{\rm
bol}$ is
\begin{equation}
\frac{L_{\rm bol}}{4.5 \times 10^{9} L_\odot} =  
\frac{\dot{\cal M}}{1 {\cal M}_\odot {\rm yr}^{-1}} ,
\label{e:lbolsfr}
\end{equation}
where the luminosity is referenced to $L_\odot= 3.83\times 10^{33}\,
{\rm erg\, s^{-1}}$, the bolometric luminosity of the sun (Allen,
1973).  The conversions of light to mass (eq.~\ref{e:mlrat}) and star
formation rate (eq.~\ref{e:lbolsfr}) are much more dependent on the
IMF parameters and adopted star formation history than the bolometric
correction (eq.~\ref{e:bcuv}), hence our mass and star formation rate
estimates should be considered indicative only.

\subsection{UV Subsamples}

(I.a) Local UV-selected starbursts are represented by the M95 dataset,
where the results are derived from HST Faint Object Camera (FOC)
images obtained with the F220W filter.  Total fluxes and effective
radii were determined for the eleven starburst regions observed in
nine galaxies (there are three detached regions in their NGC~3690
image).  The sample is at sufficiently small redshift that $k$
corrections are unnecessary. The Galactic foreground extinction is
taken from Burstein \&\ Heiles (1984), and removed from the UV
following the extinction curve of Seaton (1979) using the algorithm
given by M95.  The $\beta$ values were measured from {\em
International Ultraviolet Explorer\/} spectra, and taken directly from
M95. 

(I.b) We have obtained new data on two starburst galaxies at $z
\approx 0.4$, QNY1:32 and SGP1:10, discovered by Boyle \etal\ (1990).
The two galaxies were observed with the HST using the FOC and the
F342W filter ($\lambda_c = 3403$\AA), which has a central wavelength
in the rest frame of the galaxies ($\lambda_c/(1+z) \approx 2380$
\AA), very similar to that of the F220W filter.  The $k$ corrections
(eq.~\ref{e:kcorr}), although small ($< 0.1$ mag), were applied.  Both
galaxies consist of three closely spaced knots aligned nearly
linearly.  The 512$\times$ 512 images were corrected for non-linearity
of the FOC and the fluxes and effective radii were obtained from
curves of growth using concentric elliptical apertures.  The apertures
were centered on the central knot and had orientations and axial
ratios set to match the outer isophotes of the galaxy.  The basic
properties of the galaxies are listed in Table~\ref{t:prgals}.

Galactic extinction, listed as $A_{\rm Gal}$ in table~\ref{t:prgals},
was estimated from the foreground \HI\ column density (Stark \etal\
1992) using the Mathis (1990) extinction curve. These galaxies do not
have $\beta$ or UV color measurements, so we have to estimate $A_{\rm
int}$ by some other manner.  We have obtained flux calibrated
optical spectra for both sources, which confirms that these
are indeed starbursts (strong narrow emission lines). The SGP1:10
spectrum was obtained by us with the KPNO 4m telescope.  Alexei
Filippenko kindly obtained the QNY1:32 spectrum for us using the Lick
3m telescope.  The ratio $F_{\rm H\gamma}/F_{\rm H\beta} = 0.46 \pm
0.11$ for SGP1:10, which is consistent with case B recombination and
no intrinsic extinction, hence we assume $A_{\rm int} = 0$.  Only one
Balmer line (\Hbeta) was well detected in the spectrum of QNY1:32.  We
assume it has the median reddening $\langle \beta - \beta_0 \rangle =
1.4$ of the rest of the UV sample, hence $A_{\rm int} = 2.2$ mag.  

(I.c-e) Very high redshift galaxies are represented by three samples.
One was selected from ground based observations, and the other two
were from the Hubble Deep Field (HDF) observations (Williams \etal,
1996).  In all three cases the galaxies were selected as probable
Lyman break systems from their broad band colors using the technique
pioneered by C.\ Steidel and collaborators. We consider only the
galaxies that have been spectroscopically confirmed, which has been
done with the Keck telescope in all cases.  The confirmation spectra
frequently show\ion{C}{4} and \ion{Si}{4} absorption features which
are commonly observed in starburst galaxies.  These arise in the winds
and photospheres of high-mass stars and/or a highly ionized
interstellar medium. The crucial size measurements come from HST
imagery with WFPC2.  The spectral slope $\beta$, is measured directly
from published broad band colors in the AB system ($m_{\rm AB} = -48.6
- 2.5\log(f_\nu)$, where $f_\nu$ is the spectral flux density per
frequency interval in units of ${\rm erg\, cm^{-2}\, s^{-1}\,
Hz^{-1}}$).  The redshifts ($z \approx 3$) of the sources put the $V$
to $I$ band observations squarely in the rest frame UV.  Details on
the individual high-$z$ subsamples are as follows:

(I.c) The original selection for this subsample is from the ground based
photometry of Steidel \&\ Hamilton (1992) and Steidel \etal\ (1995).
Spectroscopic redshifts are given by Steidel \etal\ (1996a), and HST
imagery is reported by Giavalisco \etal\ (1996).  There are seven
spectroscopically confirmed objects in the sample.  Giavalisco \etal\
show that two of the seven galaxies are double.  Here we consider the
components separately assuming they have a common redshift, so the
total size of this subsample is nine.  We derive $\beta$ from the
$({\cal G - R})_{\rm AB}$ colors given by Steidel \etal\ (1996a):
\begin{equation}
\beta = 2.55({\cal G - R})_{\rm AB} - 2,
\end{equation}  
We take $R_e = R_{1/2}^T$ from Table~2 of Giavalisco \etal\ (1996).

(I.d) The first HDF sample is that of Steidel \etal\ (1996b).  Five
confirming redshifts were obtained, one of which corresponds to a
double source; thus the total subsample size is six.  Effective radii
$R_e = R_{1/2}^T$ were taken from their Table 2.  Rest-frame UV fluxes
were derived from the ${\cal R}$ band (combined F606W and F814W light)
magnitudes in their Table~1. Finally, $\beta$ was estimated from the
$(m_{\rm F606W} - m_{\rm F814W})_{\rm AB}$ colors given in the HDF
version 2 catalog (Williams \etal, 1996):
\begin{equation}\label{e:beta_vi}
\beta = 3.23(m_{\rm F606W} - m_{\rm F814W})_{\rm AB} - 2.
\end{equation}
We used $(m_{\rm F606W} - m_{\rm F814W})_{\rm AB}$ in preference to
$(m_{\rm F450W} - {\cal R})_{\rm AB}$ listed by Steidel \etal, because
the blue band in the latter may be affected by opacity from the
Ly$\alpha$ forest (Madau \etal, 1996).  

(I.e) The other HDF subsample has confirmation spectra from the DEEP
(``Deep Extragalactic Evolutionary Probe'') program (Lowenthal \etal\
1997).  A total of 11 high-$z$ detections were made for sources
not in sample I.d. Four of these belong to double sources.  Hence the
total subsample size is 15.  Effective radii were taken from their
Table 2 (listed as $r_{1/2}$), and $\beta$ was derived from the $(m_{\rm
F606W} - m_{\rm F814W})_{\rm AB}$ colors given in their Table 1 using
eq.~\ref{e:beta_vi}. 

\subsection{UV results}

\begin{figure}
\centerline{\hbox{\psfig{figure=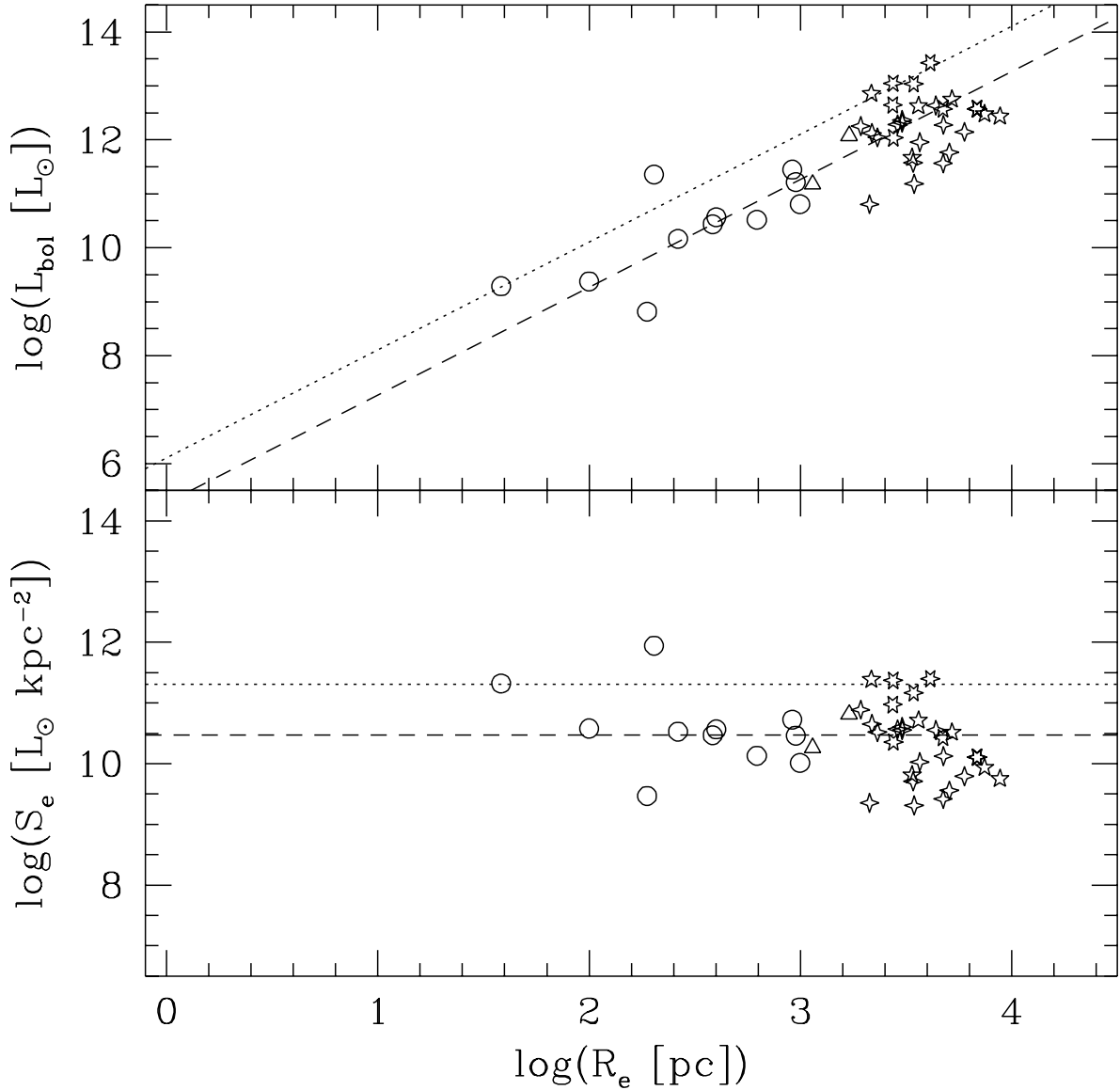,width=17cm}}}
\caption{Bolometric luminosity $L_{\rm bol}$ and
effective surface brightness $S_e$ plotted against effective radius
$R_e$ for UV-selected starbursts.  The dotted and dashed lines
correspond to $S_{e,90}$ and $S_{e,50}$ of the combined sample.  The
correspondence between symbols and subsamples is as follows: circles -
I.a.\ local starbursts (M95); triangles - I.b.\ moderate-$z$
starbursts (this work); six pointed stars - I.c.\ high-$z$ starbursts
(Giavalisco \etal\ 1996; Steidel \etal, 1996a); five pointed stars -
I.d.\ HDF high-$z$ galaxies Steidel \etal\ (1996b) sample; four
pointed stars - I.e.\ HDF high-$z$ galaxies DEEP sample (Lowenthal
\etal, 1997).  \label{f:limuv}}
\end{figure}

\begin{figure}
\centerline{\hbox{\psfig{figure=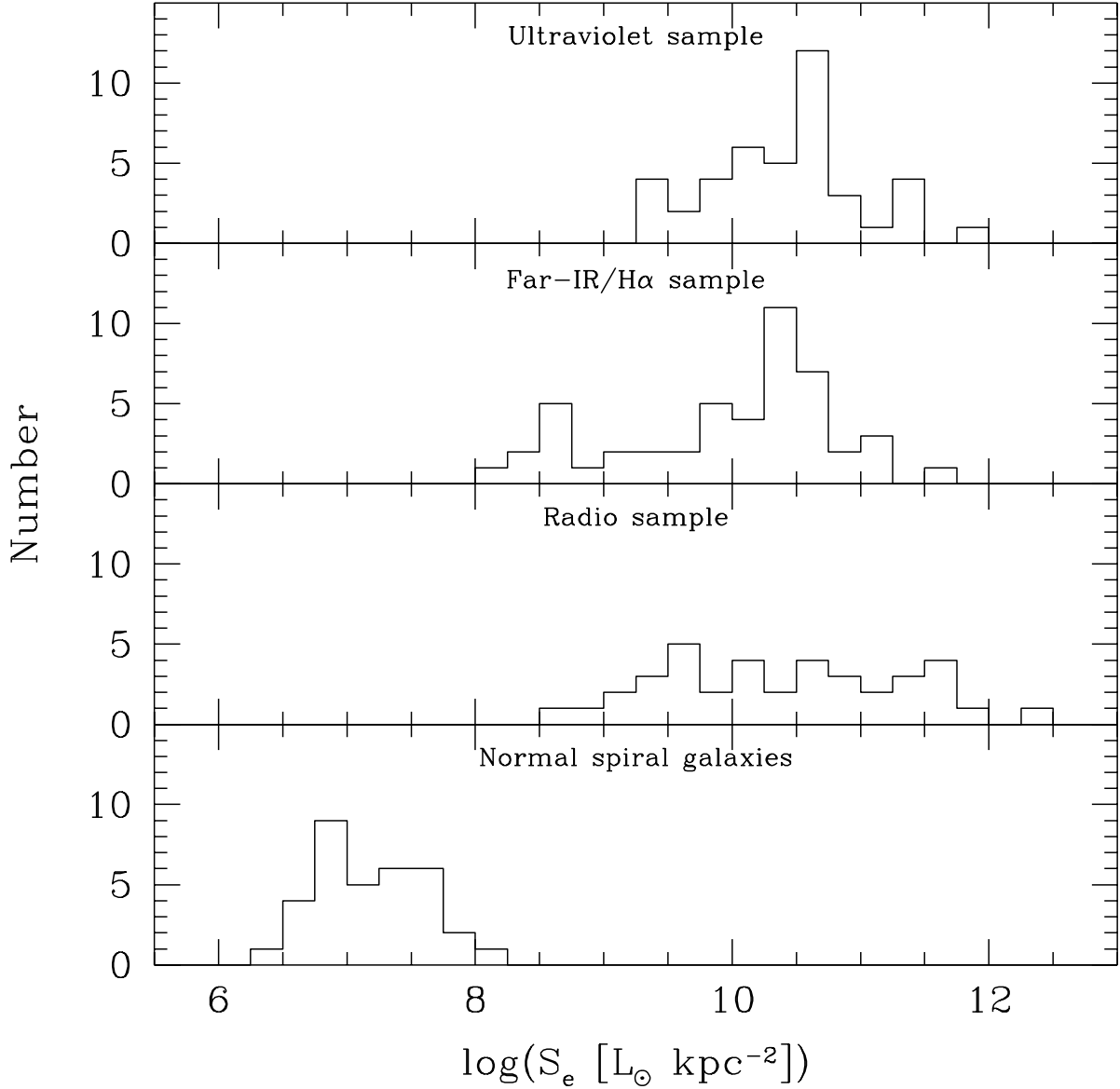,width=17cm}}}
\caption{Bolometric surface brightness distributions of
the UV, FIR/\Halpha, and radio samples compared to that derived from
\Halpha\ observations of normal galaxies 
(Ryder \&\ Dopita, 1993).\label{f:dist}}
\end{figure}

Figure~\ref{f:limuv} shows $L_{\rm bol}$ and $S_e$ as a function $R_e$
of the UV-selected samples.  The $S_{e,50}$ and $S_{e,90}$ surface
brightness levels of the combined sample are plotted as dashed and
dotted lines respectively. Figure~\ref{f:dist} shows the surface
brightness distribution of the combined UV sample in the top panel.
The combined sample has $S_{e,90} = 2.0 \times 10^{11}\, L_\odot\,
{\rm kpc^{-2}}$.  Table~\ref{t:stats} tabulates $\log(S_{e,50})$ and
$\log(S_{e,90})$ levels of the combined UV sample, and its various
subsamples.

It is immediately apparent from Fig.~\ref{f:limuv} that $S_e$ shows
little or no dependence on $R_e$ over about two orders of magnitude in
size; hence there is no dependence on $L_{\rm bol}$ over about four
orders of magnitude in luminosity.  Table~\ref{t:stats} indicates that
$S_{e,90}$ is the same within 0.1 dex for the entire sample and the
combined subsamples I.c-e containing just high-$z$ galaxies.  This is
quite small compared to the cosmological dimming (which has been
removed) of $\log(z+1)^4 = 2.4$ dex at $z=3$.

All of the starbursts in this sample are well resolved; there are no
upper limits on size, even though the samples were not biased against
point sources. The range of angular sizes of the galaxies in the
samples spans nearly two orders of magnitude similar to the range of
absolute sizes.  We checked how our results are affected by whether
{\em detached\/} substructures are counted separately or not by
considering the two galaxies in subsample I.b.  When their knots are
considered separately the brightest in each has $S_e$ about 0.25 dex
higher than the galaxy considered as a single unit.  However, the
average $S_e$ of the individual knots is close to the $S_e$ value of
the galaxy as a whole.  This indicates that the detached substructures
have a high covering factor.  We conclude that the $S_e$ measurements
are robust to within $\sim 0.3$ dex with respect to resolution
effects.

\begin{figure}
\centerline{\hbox{\psfig{figure=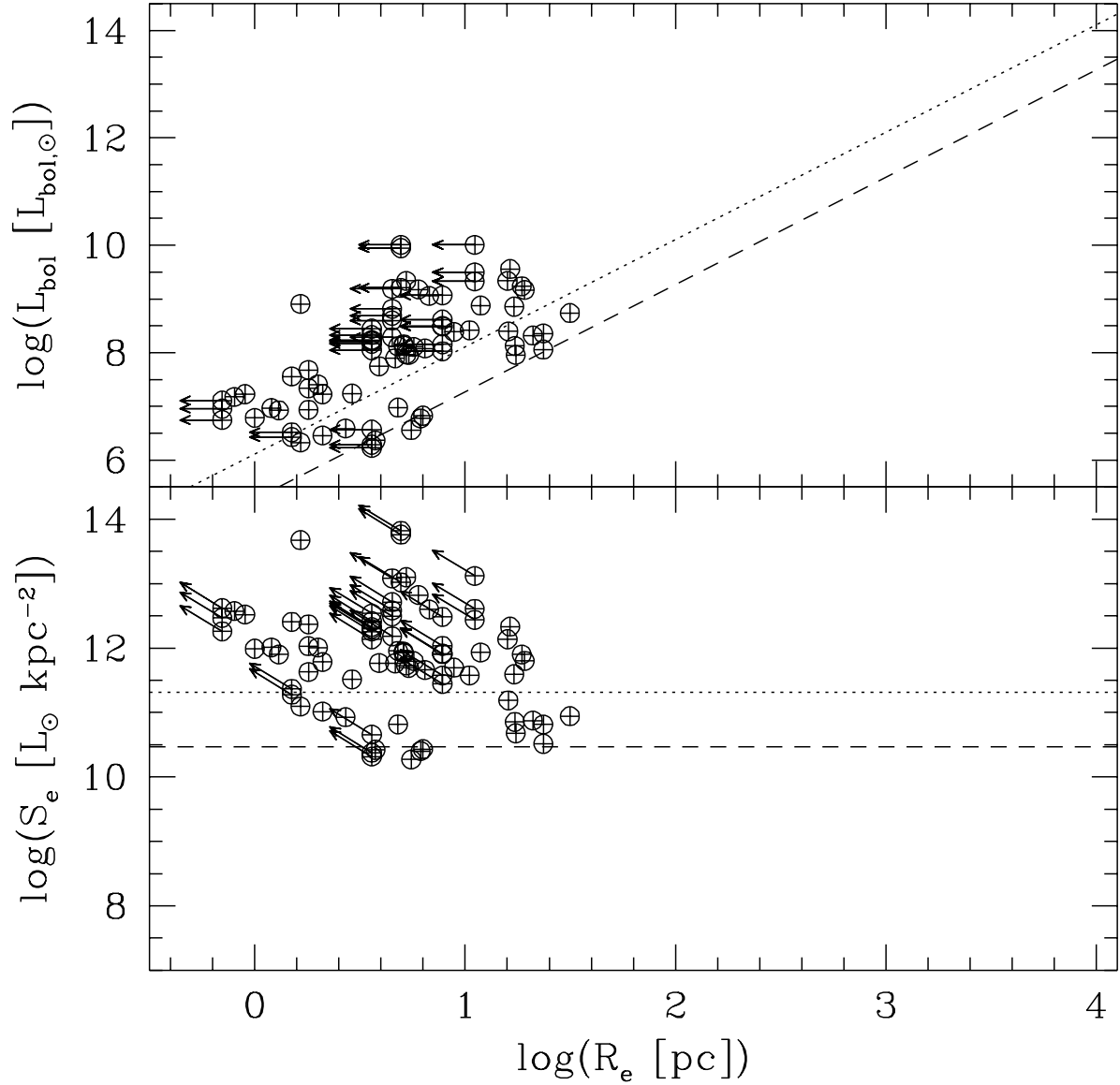,width=17cm}}}
\caption{Same as Fig.~{3} for star clusters within
starbursts (from M95).  Arrows indicate upper limits to $R_e$.  The
dotted and dashed lines have the same position as in
Fig.~{3}. \label{f:limcl}}
\end{figure}

The same can not be said about the {\em embedded\/} structure.  In
particular the star clusters within starbursts have much higher $S_e$
values than starbursts as a whole. Figure~\ref{f:limcl} shows $L_{\rm
bol}$ and $S_e$ of the star clusters with size estimates from M95
(their table 10).  The bolometric corrections were estimated using
eq.~\ref{e:bcuv}.  The $S_{e,50}$ and $S_{e,90}$ lines from
Fig.~\ref{f:limuv} are plotted for comparison.  All of these clusters
have surface brightnesses above the $S_{e,50}$ and most are above the
$S_{e,90}$ level derived from integrated starbursts.  This is
partially a selection effect since the clusters are all embedded
within the dominant diffuse background of a starburst.  In this
figure there are many upper limits to $R_e$, and thus lower limits to
$S_e$. In addition the $R_e$ estimates for the clusters in the more
distant galaxies are probably inflated (and $S_e$ underestimated) as
noted by M95.  The most intense emission from a resolved cluster comes
from NGC1705-1 with $S_e = 4.7 \times 10^{13}\, L_\odot\, {\rm
kpc^{-2}}$.  This is over two orders of magnitude more intense than
$S_{e,90}$ of the UV sample, yet NGC1705-1 has probably already faded
by a factor of $\sim 6$ from when it was an ionizing cluster (M95,
Meurer \etal, 1992).  Figure~\ref{f:limcl} demonstrates that high
$S_e$ sources can be detected in the UV, they just do not have large
scale sizes, nor do they dominate the integrated light of starbursts.

\section{Far-infrared - \Halpha\ sample}\label{s:oirsamp}

\subsection{Subsamples, method}\label{s:irmeth}

The two subsamples of FIRGs are (II.f) the sample of Armus, Heckman
\&\ Miley (1990); and (II.g) the sample of Lehnert \&\ Heckman (1995,
1996).  They were drawn from the {\em IRAS\/} catalog using the same
FIR flux limits used to create the {\em IRAS\/} Bright Galaxy Survey
(BGS; Soifer \etal, 1987, 1989).  In addition, these samples have an
infrared color selection, $f_{60}/f_{100} \geq 0.77, 0.4$,
respectively, where $f_{60}$, $f_{100}$ are the 60\micron\ and
100\micron\ flux densities from the {\em IRAS\/} Point Source Catalog
(1988). In order to minimize contamination of the samples by galaxies
dominated by active galactic nuclei (AGN) we excluded galaxies with a LINER
or Seyfert spectral classification in Veilleux \etal\ (1995).

The FIR luminosity is derived from the far-infrared flux given by
\begin{equation}
F_{\rm FIR} = 1.26 \times 10^{-11} (2.58f_{60} + f_{100})\, {\rm erg\,
cm^{-2}\, s^{-1}}, 
\end{equation}
and the units for $f_{60}$, $f_{100}$ are Jy (Helou \etal, 1985).
Almost all of the galaxies in these samples have have $F_{\rm FIR} >
F_{\rm optical}$.  Hence to a good approximation the bolometric
luminosity is equivalent to that produced in the FIR.  We
take 
\begin{equation}
F_{\rm bol} = c(f_{60}/f_{100}) F_{\rm FIR}, \label{e:bcfir}
\end{equation} 
where $c(f_{60}/f_{100})$ is the bolometric correction to $F_{\rm
FIR}$ to obtain the total FIR luminosity (wavelength range
1-1000 \micron).  The correction assumes a single dust temperature and
emissivity $\propto \nu^{-1}$ and is tabulated by Helou \etal\ (1988).
For subsample II.f we assume a single average $c(f_{60}/f_{100}) =
1.5$, on the basis of the $f_{60}/f_{100}$ color range of the sample.
For subsample II.g we use the galaxy by galaxy $c(f_{60}/f_{100})$
corrected fluxes given by Lehnert and Heckman (1996).  Since the
$L_{\rm bol}$ measurements are made in the FIR, no extinction
corrections are necessary. For the size of the starburst we take $R_e
= R_e({\rm H}\alpha)$, the \Halpha\ effective radius.

The \Halpha\ images used to estimate $R_e$ were obtained from the
ground in FWHM seeing typically 1-2\as.  Therefore we considered
measurements with $R_e < 1.0''$ to be upper limits.  There is only one
galaxy in the two subsamples with such an upper limit: 01217+0122 with
$R_e \leq 0.86''$ from subsample II.f.  Its actual $S_e$ value does
not significantly affect our results because its limiting $S_e \geq
10^{11}\, {\rm L_\odot\, kpc^{-2}}$ is already greater than $S_{e,90}$
of the combined FIR/\Halpha\ sample (see below).

\subsection{FIR/H$\alpha$ results}

Figure~\ref{f:limir} shows $L_{\rm bol}$ and $S_e$ as a function of
$R_e$ for sample II. In comparison to the UV sample, shown in
Fig.~\ref{f:limuv}, the FIR/\Halpha\ sample looks less like a constant
surface brightness correlation. The FIR/\Halpha\ sample is richer in
low surface brightness galaxies than the UV sample, as can be seen
from the surface brightness distribution shown in the second panel of
Fig.~\ref{f:dist}.  However, there is little apparent variation in the
upper envelope of $S_e$ points in this sample.  This illustrates
nicely that we are dealing with a surface brightness limit and not a
constant surface brightness correlation.  

The combined FIR/\Halpha\ sample has $S_{e,90} = 0.83
\times 10^{11}\, L_\odot\, {\rm kpc^{-2}}$.  Table~\ref{t:stats}
tabulates the $\log(S_{e,50})$ and $\log(S_{e,90})$ values for the two
subsamples and the combined FIR/\Halpha\ sample.  These values for the
combined sample are shown as dashed and dotted lines respectively in
Fig.~\ref{f:limir}.

\section{Radio sample}\label{s:radsamp}

\subsection{Sample, method}

In the radio we use the sample of Condon \etal\ (1990; hereafter C90)
which was originally drawn from the {\em IRAS\/} BGS.  The C90 sample
includes over 300 {\em IRAS\/} galaxies of all types observed at 21cm
with the Very Large Array.  The editing steps we apply, which are
discussed below, reduce this to a final sample of 38 sources.

To determine the bolometric flux we use the FIR to radio correlation.
This well-known correlation (see Helou, 1991, for a review) between
fluxes in the two wavelength regimes holds for over four decades in
luminosity with an intrinsic scatter of only 0.2 dex making it the
tightest correlation in global fluxes of galaxies.  Although the
physical explanation of the correlation is only now being addressed
(e.g.\ Lisenfeld \etal, 1996), it seems to work remarkably well for
star formation in all sorts of environments including normal disk
galaxies, starburst systems, and even residual star formation in
elliptical galaxies.  The form of the FIR - radio correlation is:
\begin{equation}
\frac{F_{\rm FIR}}{3.75 \times 10^{12}\, {\rm Hz}} = 10^q f_{\rm \nu}({\rm
21 cm})\label{e:firrad}
\end{equation}
where $q \approx 2.35$ for starburst and normal galaxies (Sanders \&
Mirabel, 1996).  We adopt this relationship to estimate $F_{\rm FIR}$
and then obtain $F_{\rm bol}$ by bolometric correcting
$F_{\rm FIR}$ using eq.~\ref{e:bcfir} and adopting $c(f_{60}/f_{100})
= 1.4$. This is appropriate for the $f_{60}/f_{100}$ ratios of our
final radio sample and corresponds to a dust temperature of $T = 45$ K
to 70 K, and a $\nu^{-1}$ emissivity (Helou \etal, 1988).

C90 report 2D-Gaussian fits to the radio images, often for a variety
of spatial resolutions and using multiple components. We derive $R_e$
from the angular effective radius taken to be
\begin{equation}
\theta_e = \frac{\sqrt{\theta_M \theta_m}}{2},
\end{equation}
where $\theta_M$ and $\theta_m$ are the major and minor axis FWHM of
the 2D fits.  For a true circularly symmetric Gaussian distribution
$\theta_e = \theta_M/2 = \theta_m/2$.  Condon \etal\ (1991) present
higher resolution $\lambda = 3.6$ cm sizes of ultraluminous FIRGs.  We
chose not to use this study because for most of the galaxies in common
to our final sample, the components they resolve only comprise
a small fraction of the 3.6 cm flux, or comparisons of the 21 cm and
3.6 cm maps suggest that much diffuse emission is missing from the 3.6
cm maps.  The few remaining galaxies have $\theta_e$ that agree to
within a factor of $\sim 2$ at 3.6 and 21 cm.

Four editing passes were done to the sample in order to use well
resolved observations and isolate the primary contributor to systems
most likely to be starbursts (i.e.\ not AGN).  Firstly, for a given
spatial resolution only components with $f_\nu({\rm 21cm})/f_\nu({\rm
21cm,total}) > 0.5$ were selected so as to reject minor contributors
to the total flux.  Secondly, for systems observed with with more than
one VLA configuration (i.e.\ at different resolutions) we selected the
best resolved observations by taking those with the largest $\theta_M
\theta_m / $ [Beam area].  Usually this meant selecting the highest
resolution observations.  Thirdly, we reject galaxies that do not obey
the radio-FIR correlation; from the IRAS and 21 cm fluxes tabulated by
C90, we calculate $q$ in eq.~\ref{e:firrad} and reject sources with $q
< 1.9$ as likely radio loud AGN (Sanders \&\ Mirabel, 1996).  Finally,
in order to exclude other known AGN, we further selected only the
systems with \HII\ type spectra in the nuclear regions according
to Veilleux \etal\ (1995).

\subsection{Radio results}

The resulting sample has a relatively narrow 1.5 dex luminosity range
centered on $L_{\rm bol} = 2\times 10^{11}\, L_\odot$. This is
primarily due to the final editing step.  The BGS subsample used by
Veilleux \etal\ (1995) is weighted towards galaxies with $L_{\rm bol}
\gtrsim 10^{11}$ (Kim \etal\ 1995), hence low luminosity galaxies were
discarded.  Rejection of AGN preferentially removes the highest
luminosity {\em IRAS\/} galaxies (Veilleux \etal, 1995; Sanders \&\
Mirabel, 1996).  The surface brightness distribution of this sample is
shown in the third panel of Fig.~\ref{f:dist}.  The $\log(S_{e,50})$
and $\log(S_{e,90})$ levels are tabulated in Table~\ref{t:stats}.  We
find $S_{e,90} = 5.1\times 10^{11}\, L_\odot\, {\rm kpc^{-2}}$ for the
radio sample, the highest of the three samples.

In order to determine the effects of the sample editing we calculated
$S_{e,50}$ and $S_{e,90}$ at the intermediate editing steps.  The most
drastic editing step was the final one, paring the sample down from
244 to 38 sources.  The sample before this step has $\log(S_{e,90}) =
11.29$, very close to that of the UV sample, but this is due to a
large number of sources with $S_e \lesssim 10^9\, {\rm L_\odot\,
kpc^{-2}}$.  These are largely removed with the \HII\ spectra
selection, but also can be removed by applying an infrared color
selection as done for sample II; if no \HII\ selection is done, and
instead only galaxies with $f_{60}/f_{100} > 0.5$ are selected, the
resulting sample is then very similar to our adopted radio sample
having $\log(S_{e,50}) = 10.09$ and $\log(S_{e,90}) = 11.64$.  The
color selection efficiently removes cool low surface brightness
emission from the radio sample, but does not discriminate against AGN,
which like intense starbursts tend to have high color temperatures.
Although we prefer to exclude AGN as best we can, application of the
$q$ editing only lowers $S_{e,90}$ by 0.04 dex. This indicates that
allowing radio loud AGN to contaminate the sample does not appear to
significantly effect the sample $S_{e,50}$ and $S_{e,90}$.

\section{Selection effects and biases}\label{s:caveats}

All samples are affected by a surface brightness selection.  The
deepest of the UV subsamples is I.e (Lowenthal \etal, 1997).  Its
limiting magnitude for spectroscopy spread over the typical Keck FWHM
$= 0.6\as$ seeing disk, corresponds to a limiting $S_e \sim 1.4 \times
10^{9}\, L_\odot\, {\rm kpc^{-2}}$ at $z = 3$, assuming the median
value of $\beta$.  Samples II and III are drawn from the BGS which has
a flux limit of $f_{60} = 5.4$ Jy, and is composed primarily of
sources unresolved by {\em IRAS\/} thus having FWHM $< 2'$.  For a
typical $f_{60}/f_{100} = 0.55$, this corresponds to $S_e \sim 3 \times
10^7\, L_\odot\, {\rm kpc^{-2}}$.  The result that none of the
galaxies in samples II and III have $S_e$ this low is due to the
infrared color selection (sample II) and \HII\ spectrum selection
(sample III).  The important point is that sample I can not reach the
same $S_e$ depth as observed in the other two samples.

The biggest systematic flux uncertainty for the UV sample comes from
how we choose to model fig.~\ref{f:uvir}.  As noted in
\S\ref{ss:dust}, we expect a $\sim 0.4$ dex in $F_{\rm FIR}/F_{220}$
from the scatter in Fig.~\ref{f:uvir}.  Indeed if we had used the
Kinney \etal\ (1994) extinction curve in our analysis we would have
derived $S_{e,50},S_{e,90}$ to be lower by 0.2 and 0.4 dex
respectively for sample I.  We have not corrected samples II and III
for the fraction of starburst light that escapes dust reprocessing.
This should be a small correction since the median $L_{\rm IR}/L_{B}
\approx 5$ for the galaxies in sample II (Lehnert \&\ Heckman 1995;
Armus \etal\ 1987).  Since much of the $B$ band flux in these galaxies
probably arises outside of the starburst we can expect $S_e$ of the
starburst to be systematically underestimated by $\lesssim 20\%$.

Quantifying the $R_e$ biases is more difficult, but we can get an
idea of the direction they will take.  Relative to our stated aim of
measuring the $R_e$ of the ionizing population, most of our techniques
overestimate $R_e$ due to the effects of the ISM.  In principle
$R_e({\rm UV})$ provides the most direct estimate of the stellar
population size.  It may be inflated by extended nebular continuum and
scattered light.  However M95 show that the former is probably small
($\sim 10\%$ in flux).  The resolution of the nearest starbursts into
stars argues against scattered light dominating in the UV.
Differential extinction (e.g.\ dust mixed in with the stars) also may
inflate $R_e({\rm UV})$ and $R_e(\Halpha)$.  Although in
\S\ref{ss:dust} we argue that the UV extinction is dominated by a
foreground screen (hence extinction will not be differential), M95
note that an idealized uniform screen can not fully describe the dust
distribution of most starbursts.  $R_e(\Halpha)$ (sample II) may be
more strongly effected by differential extinction than $R_e({\rm UV})$
(sample I), despite the sense of the wavelength difference, because
the {\em IRAS\/} galaxies are selected to be dusty.  In addition,
sample II.g is selected by edge-on appearance. Hence, \Halpha\
emission may be extincted by cool outer disk dust, far from the
starburst, resulting in a large $R_e(\Halpha)$ bias.  We may also
expect that $R_e(\Halpha)$ to be inflated in galaxies displaying
galactic winds.  However it is not clear that this is a large effect.
On the one hand, Marlowe (1997) find that the effect can be a large
(factor of five in $R_e$) in windy blue compact dwarves.  On the other
hand, using Lehnert's \&\ Heckman's (1995) statistic $\log({\rm
R}_{{\rm H}\alpha}/{\rm R}_R)$ we find no noticeable increase of
$S_{e,90}$ for subsample II.g when the ten galaxies with the strongest
winds are excluded from the subsample. As for sample III, the
$R_e({\rm 21cm})$ measurements will be least effected by extinction.
A bias towards high surface brightness will result from using the
Gaussian HWHM to approximate $R_e({\rm 21cm})$, instead of doing a
full curve of growth analysis, and by using radio interferometric
measurements preferentially of high angular resolution.  Both
procedures discard low surface brightness extended emission.  On the
other hand, radio emission tends to be more extended than star
formation in nearby galaxies (e.g.\ Marsh \&\ Helou, 1995), probably
due to the propagation of cosmic rays in the hosts.

In summary, selection effects and measurement biases may
significantly effect the $S_e$ distributions.  $R_e({\rm UV})$ and
$R_e({\rm \Halpha})$ are both likely to be inflated with respect to
the $R_e$ of the ionizing population; the latter is likely to be more
severely affected.  Competing physical and measurement biases may
cause $R_e({\rm 21cm})$ to be either under or overestimated. We expect
systematic biases up to a factor of three in UV flux.  Agreement of
$S_{e,90}$ values to within this factor indicate consistent starburst
intensity limits as best we can determine.

\section{Discussion}\label{s:disc}

\subsection{The Surface Brightness Limit}

Table~\ref{t:stats} shows that the (full) samples II and III have the
largest difference in $S_{e,90}$, 0.8 dex.  Since they were originally
drawn from the same source, the {\em IRAS\/} BGS, the factor of six
difference in $S_{e,90}$ is illustrative of the effect that details of
final sample selection and measurement bias can have.  Since the
fluxes used to calculate $S_e$ are essentially the same (tied to
$F_{\rm FIR}$), the differences in $S_{e,90}$ probably arise in the
$R_e$ estimates, perhaps for the reasons discussed in
\S\ref{s:caveats}.  The $S_{e,90}$ of sample I splits the
difference of the other two samples, agreeing with each to within a
factor of three.  This is a reasonable level of agreement as discussed
above.

We conclude that to better than an order of magnitude, the maximum
surface brightness $S_{e,90}$ of starbursts are the same, when deduced
from radio, optical and UV observations.  We find a mean 
$S_{e,90} \approx 2.0 \times 10^{11}\, L_\odot\, {\rm kpc^{-2}}$ (with
a factor of three uncertainty) as the characteristic surface
brightness of starbursts, found by averaging in the log the results of
the three samples. The lack of a simple wavelength dependence suggests
that this is a star formation intensity limit, not an opacity effect.
Starbursts orders of magnitude more intense than deduced from UV
observations are not common.

Further evidence of the physical significance of $S_{e,90}$ is given
by the central pressure $P_0$ of windy FIRGs.  For a sample of 12
galaxies (mostly starbursts) undergoing strong galactic wind Heckman
\etal\ (1990) estimate $P_0$ using the [\ion{S}{2}]$\lambda
\lambda$6716/6731 \AA\ emission line ratio.  They find the
mean ($\pm$ dispersion) $P_0 = 2.8 \pm 1.2 \times 10^{-9}\, {\rm dy\,
cm^{-2}}$.  In the starburst models of Chevalier \&\ Clegg (1985),
$P_0$ of a free flowing wind is given by $P_0 = 0.118 \dot{p}R^{-2}$,
where $\dot{p}$ is the momentum flux from the starburst, and $R$ is
the starburst radius, and the constant is for cgs units.  For our
standard starburst population model (\S\ref{s:uvmeth}) the momentum
flux due to supernovae and stellar winds is given by
\begin{equation}
\dot{p} = 2.1 \times 10^{23} \frac{L_{\rm bol}}{L_\odot}\, [{\rm erg\, cm^{-1}}].
\end{equation}
The constant varies by a factor of 1.6 for burst durations in the
range of 5 to 100 Myr.  The star formation intensity required to
produce $P_0$ is then 
\begin{equation}
\frac{S_e}{10^{11}\, L_\odot\, {\rm kpc^{-2}}} 
 = \frac{P_0}{1.63 \times 10^{-9}\, {\rm dy\, cm^{-2}}}.
\end{equation}
Hence the mean value of $P_0$ measured by Heckman \etal\ (1990)
corresponds to $S_e = 1.7 \times 10^{11}\, L_\odot\, {\rm kpc^{2}}$.
This is very close to the adopted $S_{e,90}$, but derived using
methods completely independent of those in \S\ref{s:uvsamp} -
\S\ref{s:radsamp}.

Since we are measuring the surface brightness of star formation
tracers, we can convert $S_{e,90}$ to an equivalent star formation
intensity ($\dot{\cal M}$ per unit area) of
\begin{equation}
\dot\Sigma_{e,90} \sim 45\, {\cal M}_\odot\, {\rm kpc^{-2}\,
yr^{-1}}\label{e:sfalim}
\end{equation}
using eq.~\ref{e:lbolsfr} (and bearing in mind the caveats of
Sec.~\ref{s:uvmeth}).  Strictly speaking the observed emission is
dominated by high-mass star formation.  If the lower mass limit of our
adopted Salpeter IMF is 5 \Msun, then $\dot\Sigma_{e,90}$ is reduced
by a factor of 5.5.

Because selection effects preferentially depopulate low $S_e$ sources,
90\%\ {\em or more\/} of starbursts have $S_e \lesssim 2.0 \times
10^{11}\, L_\odot\, {\rm kpc^{-2}}$.  Thus it is close to an upper
limit to the surface brightness of starbursts.  This implies that
starbursts with $S_e \approx S_{e,90}$ can only be more luminous by
being larger.  The UV results in particular show that the same limit
applies over two orders of magnitude in $R_e$ (four in $L$) and
displays little evolution out to $z \approx 3$.

The star formation intensity is much higher in starbursts relative to that
integrated over typical galactic disks.  This is illustrated in the
bottom panel of Fig.~\ref{f:dist} which shows the $S_e$ distribution
of normal disk galaxies, derived from the \Halpha\ imaging sample of
Ryder \&\ Dopita (1993). Here bolometric fluxes due to high-mass star
formation were derived from \Halpha\ fluxes using the Zanstra method.
$R_e$ estimates were kindly provided by S.D.\ Ryder, and correspond to
the effective radius of the elliptical aperture containing half of the
total \Halpha\ emission. Table~\ref{t:stats} lists $S_{e,50}$, and
$S_{e,90}$ for this sample.  The typical $S_e$ of star formation in
normal spiral galaxies is about three orders of magnitudes fainter
than that of commonly selected starbursts.  Much of the difference can
be ascribed to the filling factor of star formation.  The \Halpha\
morphology of normal spiral galaxies is characterized by numerous, but
widely separated \HII\ regions, whereas starburst galaxies are
typically dominated by a single central super-bright star forming complex.
Individual giant \HII\ regions in normal galaxies may have $S_e$
approaching that found in starbursts.

\subsection{Starburst Regulation}\label{s:klaw}

What are the regulating mechanisms that limits the global starburst
intensity?  The central pressure estimates discussed above suggests
that feedback from high-mass star formation may be involved in the
regulation (cf.\ the discussion of M82 by Lehnert \&\ Heckman 1996).
However, it is unclear how this would regulate star formation
intensity.  The near equivalence between $R_e(\Halpha)$ and the
rotation curve turn-over radius $R_{\rm to}$ found by Lehnert \&\
Heckman (1996) suggests that the global dynamics of the host in
limiting at least the size of the starburst.  Here we consider whether
the $S_e$ limit is related to the stability of inner disks of
galaxies.

Following on the work of Toomre (1964) and Quirk (1972), Kennicutt
(1989) showed that the radial gas surface density profiles
$\Sigma_g(R)$ in spiral galaxies closely follows, but is somewhat
under, the surface density at which gas disks would be unstable to
their own self gravity.  Specifically, Kennicutt defines the critical
density for star formation as
\begin{equation}
\Sigma_c = \frac{\alpha\kappa\sigma}{3.36 G}
\end{equation}
where $\sigma$ is the velocity dispersion of the gas, $\alpha = 0.67$ is an
empirically determined constant, and
\begin{equation}
\kappa = \left(R\frac{d\Omega^2}{dR} + 4\Omega^2\right)^{1/2}
\end{equation}
is the epicyclic frequency for angular frequency $\Omega = V(R)/R$ with
$V(R)$ being the rotation curve.  He found that massive star formation
in \HII\ regions seems to be inhibited except where $\Sigma_g >
\Sigma_c$.  The so called ``Kennicutt law'' has proven to be very
successful in determining the location of \HII\ regions (or lack of
them) in a variety of galaxies including the normal spirals, low
surface brightness galaxies (van der Hulst \etal, 1993), and \HII\
galaxies (Taylor \etal, 1994). The overall success of the Kennicutt
law (however cf.\ Ferguson \etal, 1996; Meurer \etal, 1996) implicates
self gravity in regulating star formation in gaseous disks.

Kennicutt's original calibration of $\alpha$ was done for galaxies
with star formation in the flat part of the rotation curve. What
happens in the rising portion of the rotation curve where starbursts
occur?  Assuming a simple linear rise, $\Omega$ is constant (solid
body rotation), and we have $\Sigma_c = \alpha \sigma
\Omega/(1.68 G)$.  We assume that $\Sigma_g \approx \Sigma_c$, since
in the Kennicutt scenario, once $\Sigma_g$ reaches or surpasses
$\Sigma_c$, star formation becomes relatively efficient, thus
regulating $\Sigma_g$.  This assumption certainly holds in particular
cases, such as NGC~3504, where $\Sigma_g$ traces $\Sigma_c$ well into
the solid body portion of the rotation curve (Kenney \etal, 1993).  On
causality grounds, the quickest time-scale expected for star formation
is the dynamical time-scale given by
\begin{equation}
t_{\rm dyn} = \sqrt{\frac{3 \pi}{16 G \rho}}
\end{equation}
(Binney \&\ Tremaine, 1987) where $\rho$ is the mean interior density.
Note that, for a cold massless disk in the potential of a dominant
spherical mass distribution, the solid body portion of the rotation
curve corresponds to a constant density core.  In that case $t_{\rm
dyn} = \case{1}{2}\pi \Omega^{-1}$.  The maximum star formation
intensity expected for gas at $\Sigma_c$ is then
\begin{equation}
\dot\Sigma_c \leq \frac{\Sigma_c}{t_{\rm dyn}} = 9\times 10^{-5}\,  
\alpha \sigma \Omega^2 ~ [{\cal M}_\odot\, {\rm yr^{-1}\,
   kpc^{-2}}] \label{e:mklaw},
\end{equation}
where $\sigma$ is in units of \kms, and $\Omega$ is in units of ${\rm
km\, s^{-1}\, kpc^{-1}}$.  This can be expressed in terms of bolometric
surface brightness using eq.~\ref{e:lbolsfr}:
\begin{equation}
S_c \approx 4.0 \times 10^5\, \alpha \sigma \Omega^2 ~ 
[L_\odot\, {\rm kpc^{-2}}] ,\label{e:klaw}
\end{equation}
using the same units as eq.~\ref{e:mklaw}.  Figure~\ref{f:klaw} plots
$S_e$ versus $\Omega$ for the galaxies in Lehnert \&\ Heckman (1996).
There is a weak correlation (correlation coefficient $R = 0.61$) in
the sense predicted by eq.~\ref{e:klaw}, which is plotted for $\alpha
= 0.67$, $\sigma = 15$ \kms\ (the central \HI\ velocity dispersion of
a representative normal disk galaxy; Dickey \etal, 1990).  More
importantly, almost all the galaxies in this sample have $S_e < S_c$
as expected.  We conclude that our results are consistent with
Kennicutt's threshold model of star formation. This implies that
starbursts may be regulated by the same gravitational instability
mechanism that applies to the outer disks of galaxies.  The
extraordinary nature of starbursts is not due to extraordinary
physics.  Rather it is due to the high densities required to reach the
instability limit, and the consequent short minimum dynamical
time-scales (see also Elmegreen, 1994).

Although application of the Kennicutt law is insightful, it does not
explain why the same surface brightness limit appears to hold over two
orders of magnitude in size.  It suggests
that the blame should be shifted to a fundamental limit on the angular
frequency $\Omega$, or to the central density since $\rho_0 \propto
\Omega^2$.  But we have no insight into whether such a limit really
exists, or why it should exist.  It also does not explain the
efficiency of star formation.  We conclude that the stability of a
gaseous disk against self gravity probably plays an important role in
limiting starburst intensity, but in and of itself is not sufficient to
explain the $S_e$ limit of starbursts.

\subsection{Galaxy formation}

How does $S_{e,90}$ relate to the global star formation history of
normal galaxies?  Can they be made in a high intensity starburst?
Normal galaxies also show preferred surface brightnesses.  Bright
spirals obey the Freeman (1970) law having exponential disks with
extrapolated {\em B\/} band central surface brightnesses of $B(0)_c =
21.65$ mag arcsec$^{-2}$, which translates to an effective surface
brightness of $S_e = 5.1 \times 10^7 L_{\rm B,\odot}\, {\rm
kpc^{-2}}$.  McGaugh \etal\ (1995) showed that this is actually an
upper limit to the surface brightness distribution of disk galaxies
when selection effects are taken into account. Elliptical (E) galaxies
fall on the so-called ``fundamental plane'' in $\log(R_e)$,
$\log(S_e)$, and $\log(\sigma)$.  Using the fundamental plane
projections of Djorgovski \&\ Davis (1987) we estimate the properties
of a representative low-luminosity E galaxy to be $R_e = 1.2$ kpc,
$\sigma = 180\, {\rm km\, s^{-1}}$, and $S_e(r_G) = 9.0 \times 10^8
L_{r_G,\odot}\, {\rm kpc^{-2}}$, whereas for a representative
high-luminosity E galaxy we adopt $R_e = 12$ kpc, $\sigma = 350\, {\rm
km\, s^{-1}}$, and $S_e(r_G) = 1.4 \times 10^8 L_{r_G,\odot}\, {\rm
kpc^{-2}}$.  Here the observations are defined in the $r_G$ passband
(Djorgovski, 1985).

An estimate of the minimum time needed to build a galaxy (e.g.\ at the
starburst intensity limit) is given by 
\begin{equation}
t_{\rm build} = \frac{(M/L) S_e}{\dot\Sigma_{e,90}}\, {\rm Myr},
\end{equation}
where $\dot\Sigma_{e,90}$ is given in eq.~\ref{e:sfalim}.  Taking
$M/L_B \approx 2.5$ to be appropriate for spirals (Puche \&\ Carignan,
1991) and $M/L_{r_G} \approx 5$ for E galaxies (Djorgovski \&\ Davis,
1987) yields $t_{\rm build} \approx 3$, 98, and 15 Myr for spirals,
low luminosity E and high luminosity E galaxies respectively.  

As before, the minimum time-scale for star formation will be on the
order of the dynamical time.  This is given by
\begin{equation}
t_{\rm dyn} \approx \frac{2 R_e}{\sigma} \approx \frac{2 R_e}{V_{\rm circ}}
\end{equation}
(following equation 4-80b in Binney \&\ Tremaine, 1987).  Taking $R_e
= 1.68 \alpha^{-1}_B$ (Freeman, 1970) and using the data of Puche \&\
Carignan (1991) typically $t_{\rm dyn} \sim 60$ Myr for spiral galaxy
disks.  Using the representative E galaxy parameters listed
above, $t_{\rm dyn} \approx 13, 67$ Myr for low and high luminosity
E galaxies respectively.

We see that $t_{\rm build}$ is on the order of, or larger than,
$t_{\rm dyn}$ for E galaxies, and so it is plausible that they were
made by a maximum intensity starburst -- this scenario would not
violate burst duration limits set by causality.  It is thought that
the epoch of E galaxy formation is at $z \gtrsim 1$ (e.g.\ Bender
\etal, 1996), hence it is plausible that the high-$z$ starbursts are
elliptical galaxies forming.  This issue is further addressed below.

Spiral disks, on the other hand, have $t_{\rm build} \ll t_{\rm dyn}$,
so a maximum intensity burst would produce many more stars in a
dynamical time-scale than are observed in spiral disks. These
considerations are in accord with observations of present day disk
galaxies which indicate that star formation occurs on a few Gyr or
longer time-scale (Kennicutt \etal, 1994).  Although maximum intensity
bursts are too strong to build disks, disk galaxies usually have
central high surface brightness bulges.  Bulges fall on the
fundamental plane and thus they can be made by maximum intensity
bursts.  Indeed, the surface brightness profiles of central starburst
galaxies are somewhat akin to those of spiral galaxies, but with the
``bulge'' being blue and corresponding to the starburst (cf.\ Schade
\etal\ 1996; Lilly \etal\ 1995).

\subsection{Extinction in high redshift galaxies}

In \S\ref{s:uvsamp} we showed that there is little apparent $z$
evolution in $S_e$ of the UV-selected sample.  One of the most
important corrections we make to the UV data is the extinction
correction which turns out to be substantial.  Here we explore whether
these corrections are reasonable, and the implications arising from
the extinction correction.

Figure~\ref{f:betadist} shows the distribution of $\beta$ values for
the low and high redshift galaxies.  Here we plot only one point in
the cases where a galaxy was separated into multiple starburst
components, since the same $\beta$ estimate was used for all
components in these cases.  The arrows show the intrinsic $\beta_0 =
-2.5$ for an unreddened ionizing population, and the median $\beta$ of
the sample.  The median shifts by less than 0.05 in $\beta$ if the
local sample is excluded.  Furthermore, the range of observed $\beta$
is virtually identical at high and low redshifts.  The high-$z$
starbursts thus have very similar UV color distributions to the low-$z$
sample.

Figure~\ref{f:betadist} illustrates that most starbursts, including
the high-$z$ ones, are redder than naked ionizing populations. The
high-$z$ samples are selected to have the bluest possible rest-frame
UV colors.  Young unreddened populations will preferentially be
selected if they exist.  Either there are very few high-$z$
galaxies dominated by ionizing populations, and/or some of the
ionizing populations are reddened.

Opacity from the Ly$\alpha$ forest may redden the photometric $\beta$
values, if the blue filter in the color index samples rest wavelengths
bluewards of 1216\AA.  This is likely to be a problem only for
subsample I.c.  Excluding this subsample changes the median $\beta$
by less than 0.05.  Hence Ly$\alpha$ forest opacity is not the
reddening mechanism.

Our interpretation is that the reddening is due to dust extinction.
We know this is valid for the local UV sample (see Fig.~\ref{f:uvir}).
Unfortunately the confirmation spectra of the high-$z$ galaxies do not
extend to the rest-frame optical where the emission lines should be
strong, allowing confirmation of the presence of an ionizing
population.  The same rest frame UV absorption features are seen in
both the low and high-$z$ samples, however they are relatively narrow
in the high-$z$ sample (Lowenthal \etal, 1997), so it is not clear
whether they have a stellar or interstellar origin.  Hence, we can not
rule out the possibility that the high-$z$ sample is contaminated by
older non-ionizing stellar populations.  It is unlikely that there is
significant contamination by faded starbursts, because aging has a
more severe effect on luminosity than our extinction corrections.  It
will take $\sim 400$ Myr after a starburst turns off for it to redden
to the observed median $\beta = -1.1$, during that time it will have
faded by a factor of $\sim 500$ (assuming the burst duration $\ll 400$
Myr; see Fig.~16 of M95), whereas the extinction correction for this
$\beta$ is a factor of eight in the F220W band.  Hence, post-bursts
would have been that much brighter when they were ionizing
populations, and such ultra-luminous starbursts are not detected at
high-$z$.  More worrisome is contamination by extended duration star
forming populations.  However, durations longer than 300 Myr are
required to reach the median $\beta$ even if the IMF slope is $\alpha
= 3.3$ (LH95).  If contamination is severe we would have the curious
effect that $S_e$ and $\beta$ evolution ``conspire'' to mimic no
evolution and a simple extinction correction. We prefer the simpler
interpretation that the same physical mechanism, extinction by dust,
reddens both the low and high-$z$ starbursts.

The top axis of Fig.~\ref{f:betadist} translates $\beta$ to extinction
at 1620\AA\ (the approximate rest wavelength of the HDF observations
longward of the Lyman Limit) using the C94 extinction law.  The median
extinction is 2.9 magnitudes at this wavelength (a factor of 15 in
flux).  For the Kinney \etal\ (1994) extinction law, the median
extinction would be 2.0 magnitudes.  In either case, the extinction is
substantial.

In their pioneering paper, Madau \etal\ (1996) use these high-$z$
galaxies to evaluate the metal enrichment history of the early
universe deriving {\em apparent\/} metal ejection rates
$\dot{\rho}_Z(z) =$ 3.6, 0.62, 1.1 $\times 10^{-4}\, {\rm M_\odot\,
yr^{-1}\, Mpc^{-1}}$ at $z = 2.75, 3.25, 4$ (see Madau \etal\ for
details).  They note that these values are underestimates because (1)
they do not sample the full luminosity function, (2) they 
represent only the least dusty systems, and (3) they assume no dust
extinction in their analysis.  We can improve on their work by
correcting their results for the median $A_{1620} = 2.9$ mag
extinction seen in starbursts.  This yields $\dot{\rho}_Z(z) = 54,
9.4, 17 \times 10^{-4}\, {\rm M_\odot\, yr^{-1}\, Mpc^{-1}}$ for the
above three redshifts.  Strictly speaking these results are still
lower limits to the true $\dot{\rho}_Z(z)$ of the universe.

Figure~\ref{f:metalprod} shows our revised version of the
$\dot{\rho}_Z(z)$ evolution diagram (cf.\ Fig.~9 of Madau \etal).
Here we adopt the flat cosmology used by Madau \etal\ $H_0 = 50\, {\rm
km\, s^{-1}\, Mpc^{-1}}$, $q_0 = 0.5$, and assume $\Lambda = 0$.  The
$x$ axis here is look-back time in linear units, and the $y$ axis is
linear in $\dot{\rho}_Z(z)$.  All data points are shown as lower
limits: the high-$z$ data because completeness corrections still have
not been addressed; and the data with $z < 1$ because they have not
been corrected for extinction.  If the extinction correction to the
other points is small, then our new estimates suggest that the epoch
of peak galaxy formation may be as early as $z \approx 3$.  Clearly
much work needs to be done to put all estimates of $\dot{\rho}_Z(z)$
on the same scale.

Our new estimates of $\dot{\rho}_Z(z)$ are well above the mean metal
ejection rate of the universe $\langle \dot{\rho}_Z \rangle \approx
4.2 \times 10^{-4}\, {\rm M_\odot\, yr^{-1}\, Mpc^{-1}}$ (Madau \etal,
1996).  Strictly speaking it too is a lower limit, since it neglects
low surface brightness galaxies and intergalactic and intracluster gas
in the estimation.  Mushotzky \&\ Loewenstein (1997) provide an
estimate of $\langle \dot{\rho}_Z \rangle = 3.4 \times 10^{-3}\, {\rm
M_\odot\, yr^{-1}\, Mpc^{-1}}$ in the redshift interval $z = 1$ to 6
-- the epoch of elliptical galaxy formation.  This level is plotted as
a dashed line in Fig.~\ref{f:metalprod}.  In Madau \etal\ it seemed
possible that the universe was relatively quiescent at high redshifts.
It now appears that the universe was forming stars, and producing
metals, at a high rate at $z > 2$. Our results imply that rather than
a quiescent universe we are observing an active but moderately
obscured early universe.

\section{Summary}\label{s:conc}

We have examined a diverse set of observations of starburst galaxies,
spanning six decades of wavelength, sizes ranging from $\sim$0.1 to 10
kpc, and redshifts out to $z = 3.5$.  The observations were used to
estimate effective radii $R_e$, total luminosities $L$, and effective
surface brightnesses $S_e$ on the bolometric scale.  The samples were
drawn from observations obtained in 1.\ the rest frame space
ultraviolet (UV), 2.\ the far-infrared and \Halpha\ (FIR/\Halpha), and
3.\ 21cm radio emission.  The UV data, which spans the largest range
in $z$, were consistently corrected for dust extinction as estimated
from the UV reddening of the starbursts.   We have demonstrated
that this correction adequately models the redistribution of radiation
from the UV to the FIR in nearby starbursts. 

The extinction in the UV is significant.  High-$z$ galaxies observed
in the rest frame UV are too red to be naked starbursts.  If this
reddening is due to dust, the typical implied extinction is $\sim$ 2
to 3 mag at $\lambda = 1620$\AA.  The corrected metal production rate
at $z \gtrsim 3$ is then $\dot\rho_Z \gtrsim 2 \times 10^{-3}\,
\Msun\, {\rm Mpc^{-3}\, yr^{-1}}$, about five times or more higher than
Madau \etal's (1996) estimate of the Hubble time averaged $\langle
\dot\rho_Z \rangle \approx 4.2 \times 10^{-4}\, \Msun\, {\rm Mpc^{-3}\,
yr^{-1}}$, but in good agreement with Mushotzky \&\ Loewenstein's
(1997) estimate of $\langle \dot\rho_Z \rangle \approx 3.4 \times
10^{-3}\, \Msun\, {\rm Mpc^{-3}\, yr^{-1}}$ during the epoch of
elliptical galaxy formation.  Depending on the amount of extinction
required for the data having $z < 2$, the peak in $\dot\rho_Z(z)$
could be pushed back to $z \gtrsim 2$.  Applying plausible extinction
corrections illustrates that rather than being quiescent, much of the
star formation in the early universe may be obscured.

The upper limit to the surface brightness of starbursts is
parameterized by $S_{e,90}$ - the 90th percentile surface brightness
of the samples.  This statistic is susceptible to systematic
uncertainties due to measurement biases and the heterogeneous sample
selection.  We find that the samples have the same $S_{e,90}$ to
within a factor of three. This is about the level of agreement we
expect for intrinsically similar samples affected by known flux
biases.  We conclude that to within a factor of a few the same
$S_{e,90}$ applies to both dusty and relatively unobscured starbursts,
it does not vary with size for $R_e \gtrsim 0.1$ kpc, and shows little
evolution out to $z \approx 3.5$.

Our adopted 90th percentile surface brightness limit is the mean value
(in the log) $S_{e,90} \approx 2.0 \times 10^{11}\, L_\odot\, {\rm
kpc^{-2}}$ of the three samples.  The absence of a strong wavelength
variation of $S_{e,90}$ indicates that it corresponds to a physical
limit on star formation intensity, rather than being an opacity
effect.  Further evidence of the physical significance of $S_{e,90}$
comes from the central ISM pressure $P_0 = 2.8 \pm 1.2 \times
10^{-9}\, {\rm dy\, cm^{-2}}$ measured in galaxies (mostly starburst)
undergoing a strong galactic wind by Heckman \etal\ (1990).  Using the
models of Chevalier \&\ Clegg (1985) and Leitherer \&\ Heckman (1995),
the corresponding surface brightness of the stellar population
pressurizing the ISM is $S_e = 1.7 \times 10^{11}\, L_\odot\, {\rm
kpc^{2}}$.  This is identical to $S_{e,90}$, but is derived from
completely different considerations.

Our adopted $S_{e,90}$ corresponds to a star formation intensity of
$\dot\Sigma_{e,90} \sim 45 {\cal M}_\odot\, {\rm kpc^{-2}\, yr^{-1}}$.
Since starbursts represent the most intense star formation in the
universe, and since selection effects depopulate low $S_e$ sources,
these values of $S_{e,90}$ and $\dot\Sigma_{e,90}$ parameterize the
maximum {\em global\/} intensity of star formation observed in the
universe. Within starbursts there are localized sources with $S_e \gg
2 \times 10^{11}\, L_\odot\, {\rm kpc^{-2}}$.  These are star
clusters.  However, their integrated light does not dominate starburst
emission, and their sizes are limited to $R_e \lesssim 10$ pc.  The
intense cluster forming mode of star formation apparently does not
operate for $R_e \gtrsim 0.1$ kpc.

Our results imply that some robust mechanism is limiting the global
intensity of starbursts.  The $P_0$ comparison implicates mechanical
energy feedback from supernovae and stellar ejecta in regulating
starbursts.  We also examined the consequences of disk stability
models (e.g.\ Kennicutt, 1989) to the central portions of galaxies
where the rotation curve is solid body like and most starbursts are
observed to occur.  There we expect an upper limit to $S_e \propto
\sigma \Omega^2 \propto \sigma \rho$, where $\sigma$ is the gas velocity
dispersion in the disk, $\Omega$ is the angular frequency, and $\rho$ is the
spherical volume density.  We show that most starbursts with rotation
curve data obey this limit for an adopted central velocity
dispersion $\sigma = 15$ \kms.  This model implies that starbursts
have higher $S_e$ than normal disk galaxies because the star formation
occurs in the center of galaxies where the critical density is high
and the star formation time-scale is short.  Unfortunately, this model
(while illuminating) does not explain the constancy of $S_{e,90}$, it
only casts it in terms of a limit on $\Omega$ or the central value of
$\rho$. 

Normal galactic disks could not have been made in a maximum intensity
burst because it would have been accomplished in much shorter than the
dynamical time scale $t_{\rm dyn}$, the minimum time set by causality
arguments.  However, bulges, which often dominate the center of disk
galaxies, can be made in a maximum intensity burst without violating
such causality limits.  The same goes for elliptical galaxies.
Considering the $R_e$ and $L$ of the high-$z$ galaxies, and the
agreement of the corrected metal production rate derived from them
with that expected at the epoch of E galaxy formation, it is tempting
to speculate that at $z \approx 3$ we are witnessing the formation of
elliptical galaxies (Giavalisco \etal\ 1995).

{\em Acknowledgements:\/} Numerous astronomers watched this work
develop and provided useful hints of samples to include, and other
suggestions.  In particular we would like to thank Mark Dickinson,
Mauro Giavalisco, Jeff Goldader, Piero Madau, Stuart Ryder, Nick
Scoville, Min Yun, and Esther Zirbel.  Asao Habe, Chris Mihos, Colin
Norman, and Rosie Wyse provided useful comments on earlier drafts of
this paper.  We are grateful to Alexei Filippenko for obtaining the
QNY1:32 spectrum for us.  Jim Condon saved us a lot of effort by
emailing us data from C90.  We are grateful for the support we
received from NASA through HST grant number GO-05491.01-93A and LTSA
grant number NAGW-3138.  This paper is in large part based on
observations with the NASA/ESA {\em Hubble Space Telescope\/} obtained
at the Space Telescope Science Institute, which is operated by the
Association of Universities for Research in Astronomy, Inc., under
NASA contract NAS5-26555. Literature searches were performed using
NED, the NASA/IPAC Extragalactic Database, a facility operated by the
Jet Propulsion Laboratory, Caltech, under contract with the National
Aeronautics and Space Administration.

% --------------------------------------------------------------------

\begin{deluxetable}{cccccccccc}
\tablenum{1}
\tablecolumns{10}
\tablewidth{0pc}
\tablecaption{FOC observations of moderate redshift starbursts 
\label{t:prgals}}
\tablehead{\colhead{Galaxy} & \colhead{$z$} & \colhead{$m_{\rm F342W}$} &
\colhead{$A_{\rm gal}$\tablenotemark{a}} & \colhead{$A_{\rm
int}$\tablenotemark{b}} & \colhead{$a_e$} & \colhead{$a/b$} & 
\colhead{$D$} & \colhead{$\log(L_{\rm bol}/L_\odot)$} & \colhead{$R_e$} \\
\colhead{} & \colhead{} & \colhead{(STMAG)} & \colhead{(mag)} & \colhead{(mag)} & 
\colhead{($''$)} & \colhead{} & \colhead{(Mpc)} & \colhead{} & 
\colhead{(kpc)}}
\startdata
QNY1:32 & 0.4424 & 19.93 & 0.16 & 2.2 & 0.32 & 2.03 & 1338 & 12.06 & 1.70 \nl
SGP1:10 & 0.4237 & 19.87 & 0.17 & 0.0 & 0.20 & 1.70 & 1294 & 11.18 & 1.14
\enddata
\tablenotetext{a}{Galactic extinction evaluated at observed 
$\lambda = 3403$\AA.}
\tablenotetext{b}{Intrinsic extinction evaluated at rest 
$\lambda_0 = 2320$\AA.}
\end{deluxetable}

% --------------------------------------------------------------------

\begin{deluxetable}{lrccccl}
\tablecolumns{7}
\tablewidth{0pc}
\tablecaption{Statistics of samples and subsamples
\label{t:stats}}
\tablehead{\colhead{(sub)sample} & \colhead{$N$} &
\colhead{$\log(S_{\rm min})$\tablenotemark{a}} & \colhead{$\log(S_{e,50})$\tablenotemark{a}} & 
\colhead{$\log(S_{e,90})$\tablenotemark{a,b}} & \colhead{$\log(S_{\rm max})$\tablenotemark{a}} &
\colhead{Description} }
\startdata
I.       &  42 &  9.31 & 10.49 & 11.31 & 11.94 & {\footnotesize Total UV sample} \nl
~~I.a    &  11 &  9.47 & 10.53 & 11.32 & 11.94 & {\footnotesize Local UV sample (M95)} \nl
~~I.b    &   2 &  9.93 & 10.10 & \nodata & 10.26 & {\footnotesize New moderate $z$ starbursts} \nl
~~I.c    &   6 &  9.75 & 10.23 & \nodata & 11.39 & {\footnotesize High-$z$, Steidel \etal\ (1996a)} \nl
~~I.d    &   8 & 10.11 & 10.70 & \nodata & 11.46 & {\footnotesize HDF, Steidel \etal\ (1996b)} \nl
~~I.e    &  15 &  9.31 & 10.12 & 10.63 & 10.89 & {\footnotesize HDF, Lowenthal \etal\ (1997)} \nl
~~I.c-e &  29 &  9.31 & 10.43 & 11.21 & 11.40 & {\footnotesize All high-$z$ UV subsamples} \nl
II.      &  48 &  8.23 & 10.24 & 10.92 & 11.50 & {\footnotesize FIR/\Halpha\ sample} \nl
~~II.f   &  18 &  9.83 & 10.52 & 10.08 & 11.51 & {\footnotesize Armus \etal\ (1990)} \nl
~~II.g   &  30 &  8.23 &  9.69 & 10.57 & 10.90 & {\footnotesize Lehnert \&\ Heckman (1995,1996)} \nl
III.     &  38 &  8.55 & 10.33 & 11.71 & 12.32 & {\footnotesize 21cm obs. Condon \etal\ (1990)} \nl
IV.      &  79 & 10.27 & $> 11.92$\tablenotemark{c} & $> 12.73$\tablenotemark{c} & $\geq 13.82$\tablenotemark{c} & {\footnotesize Star clusters in the UV (M95)} \nl
V.       &  34 &  7.37 &  7.22 &  7.69 &  8.22 & {\footnotesize Normal disks: Ryder \&\ Dopita (1993)} 
\enddata
\tablenotetext{a}{$S_e$ units are $L_\odot\, {\rm kpc^{-2}}$}
\tablenotetext{b}{Calculated only if (sub)sample size is $N \geq 10$.}
\tablenotetext{c}{lower limits due to large number of $R_e$ upper limits.}
\end{deluxetable}
% 
%------------------------------------------------------------------

\onecolumn

% \pino

\begin{figure}
\centerline{\hbox{\psfig{figure=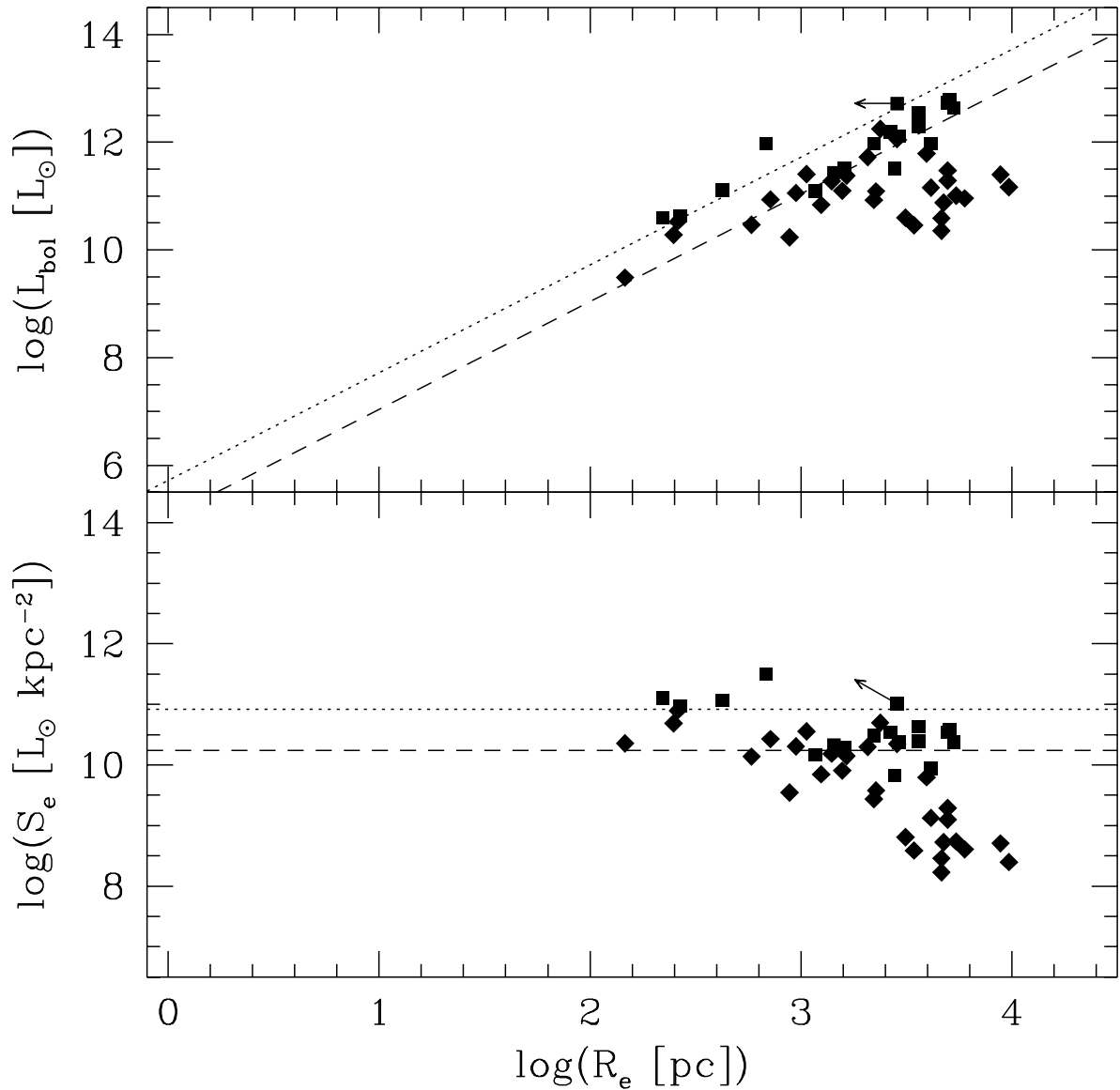,width=17cm}}}
\caption{Same as as Fig.~{3} for the FIR/\Halpha\
sample, with the dotted and dashed lines showing $S_{e,90}$ and
$S_{e,50}$ limits of this sample. Symbol - subsample correspondence:
squares - II.f.\ (Armus \etal\ 1990); diamonds - II.g.\ (Lehnert \&\
Heckman, 1995, 1996).
\label{f:limir}}
\end{figure}

\begin{figure}
\centerline{\hbox{\psfig{figure=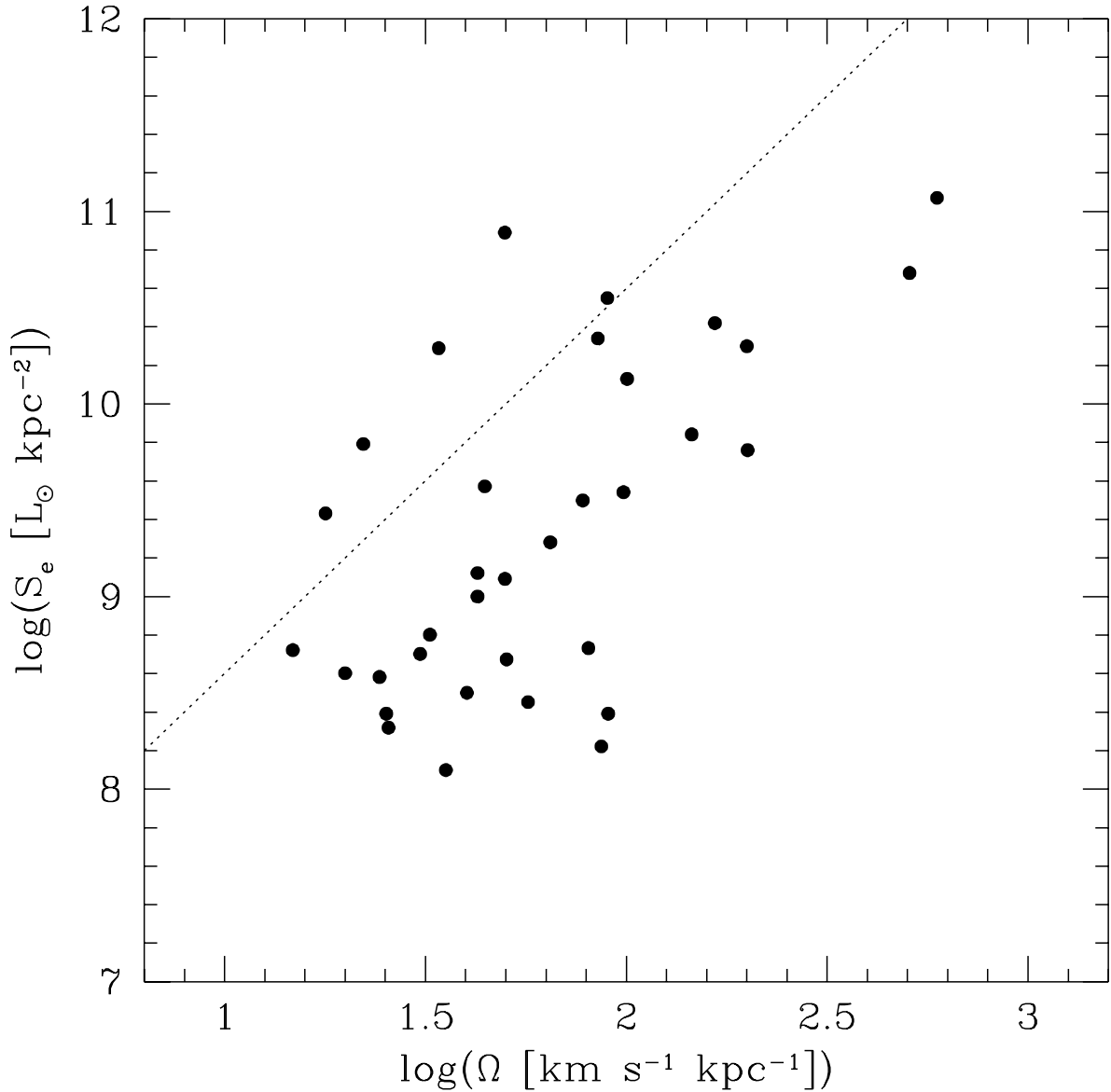,width=17cm}}}
\caption{Bolometric surface brightness versus
angular frequency $\Omega$ in the rising portion of the rotation curve
for galaxies in subsample II.g (Lehnert \&\ Heckman 1996). Here
inclinations were estimated from the axial ratios using $\cos{i}
\approx b/a$.  The dotted line shows the expected upper limit expected
for gas at $\Sigma_c$ forming stars over a dynamical time-scale.
Since full data are missing for much of subsample II.g we supplemented
their kinematics data with published rotation curve data, and have
taken $R_e = R_{\rm to}$ (the rotation curve turnover radius) for
galaxies with missing $R_e$ (required to estimate $S_e$). The
supplemental data came from the following sources: Armus \etal\
(1990), Carozzi (1977), Carozzi-Meyssonnier (1978), Durret \&\
Bergeron (1988), O'Connell \&\ Mangano (1978), Peterson (1980), Rubin
\etal (1982), and Zhang \etal\ (1993).\label{f:klaw}}
\end{figure}

\begin{figure}
\centerline{\hbox{\psfig{figure=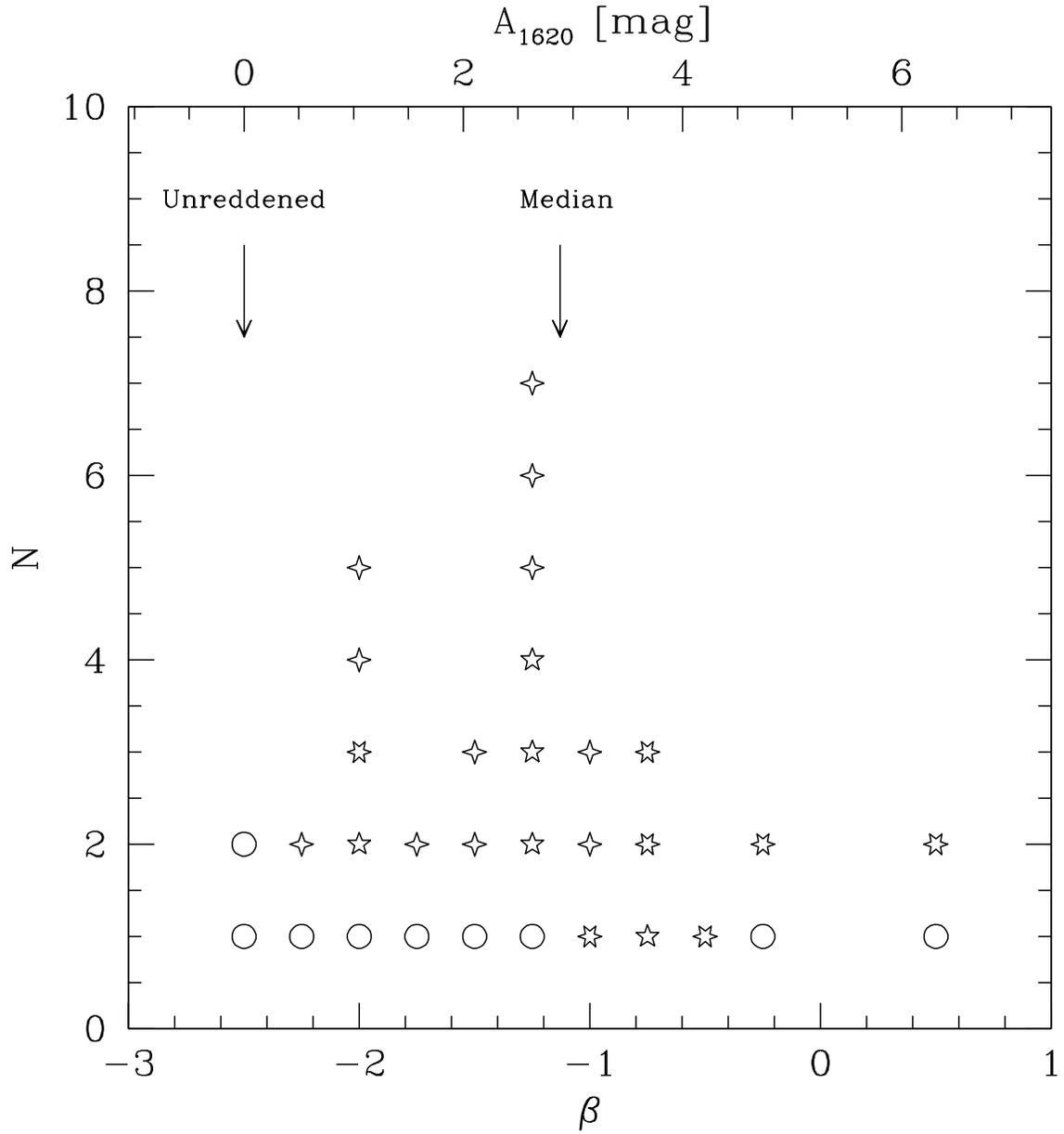,width=17cm}}}
\caption{Distribution of UV spectral slopes $\beta$
for the UV sample.  The symbols are the same as for Fig.~{3}. Only one
point is plotted where a galaxy resolves into multiple starburst
knots.\label{f:betadist}}
\end{figure}

\begin{figure}
\centerline{\hbox{\psfig{figure=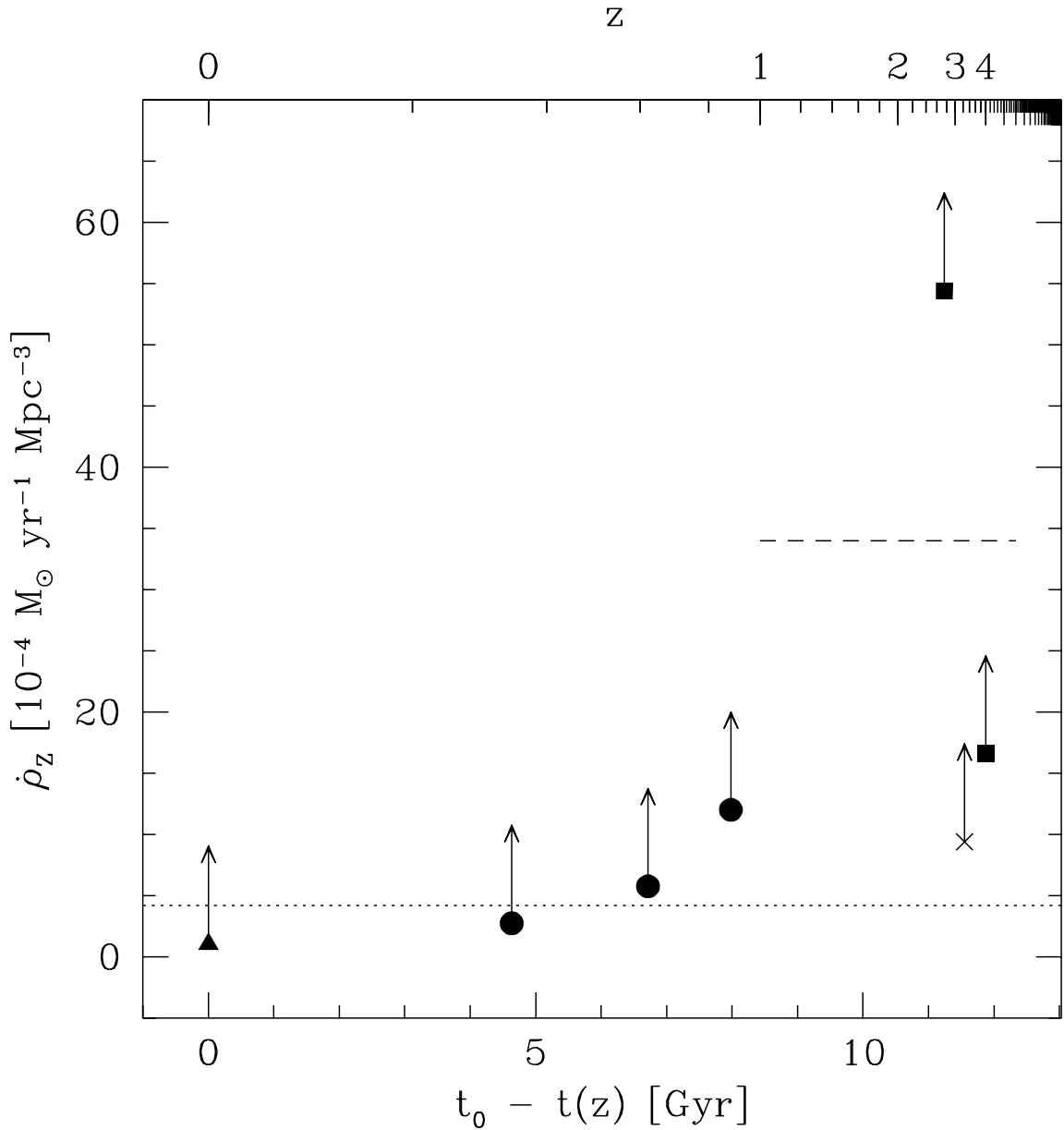,width=17cm}}}
\caption{Evolution of metal production rate as a
function of look-back time (bottom axis), and redshift (top axis). The
symbol correspondence is as follows: squares - Madau \etal\ (1996); x
- Steidel \etal\ (1996a); circles - Lilly \etal\ (1996); triangle -
Gallego \etal\ (1995).  The data points are all lower limits.  For $z
> 2$ this is because completeness corrections have not been made.  For
$z < 1$ this is because no dust extinction corrections have been
made. The dotted line shows the Hubble time averaged metal production
rate estimated by Madau \etal\ (1996).  The dashed line segment shows
the mean metal production rate during the epoch of elliptical galaxy
formation as estimated by Mushotzky \&\ Loewenstein (1997). 
\label{f:metalprod}}
\end{figure}

\end{document}